\documentclass[longauth]{aa}

\usepackage{graphicx}
\usepackage[table]{xcolor}
\usepackage{placeins}
\usepackage{natbib}
\usepackage{scalerel}
\usepackage{comment}
\usepackage[utf8]{inputenc}
\usepackage{sidecap}
\usepackage[switch, modulo]{lineno}
              
\makeatletter
\renewcommand*\aa@pageof{, page \thepage{} of \pageref*{LastPage}}
\makeatother

\renewcommand{\linenumbers}[0]{}

\usepackage{txfonts}
\usepackage[pdfencoding=auto,psdextra]{hyperref}
\hypersetup{
    colorlinks=true,
    linkcolor=blue,
    filecolor=magenta,      
    urlcolor=blue,
    citecolor=blue
}
\usepackage{xspace}
\usepackage{euclid}

\newcommand*{\Zoobot}{\texttt{Zoobot}\xspace}

\begin{document} 

\defcitealias{Q1}{Q1}

   \title{\Euclid: Quick Data Release (Q1) -- Exploring the detailed visual morphology of galaxies in clusters\thanks{This paper is published on behalf of the Euclid Consortium.}}

   \authorrunning{P. Mas-Buitrago et al.}
   \titlerunning{Detailed visual galaxy morphology in clusters}

\newcommand{\orcid}[1]{} 		   
\author{P.~Mas-Buitrago\orcid{0000-0001-8055-7949}\thanks{\email{pmas@cab.inta-csic.es}}\inst{\ref{aff1}}
\and S.~Kruk\orcid{0000-0001-8010-8879}\inst{\ref{aff2}}
\and D.~O'Ryan\orcid{0000-0003-1217-4617}\inst{\ref{aff2}}
\and M.~T.~Nardone\orcid{0009-0001-4102-9630}\inst{\ref{aff2}}
\and A.~La~Marca\orcid{0000-0002-7217-5120}\inst{\ref{aff3},\ref{aff4}}
\and P.~Awad\orcid{0000-0002-0428-849X}\inst{\ref{aff4}}
\and M.~Baes\orcid{0000-0002-3930-2757}\inst{\ref{aff5}}
\and M.~Fossati\orcid{0000-0002-9043-8764}\inst{\ref{aff6},\ref{aff7}}
\and T.~G\'eron\orcid{0000-0002-6851-9613}\inst{\ref{aff8}}
\and Z.~Ghaffari\orcid{0000-0002-6467-8078}\inst{\ref{aff9},\ref{aff10}}
\and M.~Girardi\orcid{0000-0003-1861-1865}\inst{\ref{aff11},\ref{aff9}}
\and C.~J.~Lintott\orcid{0000-0001-5578-359X}\inst{\ref{aff12}}
\and J.~G.~Sorce\orcid{0000-0002-2307-2432}\inst{\ref{aff13},\ref{aff14}}
\and B.~Altieri\orcid{0000-0003-3936-0284}\inst{\ref{aff2}}
\and F.~Durret\orcid{0000-0002-6991-4578}\inst{\ref{aff15}}
\and V.~Pettorino\orcid{0000-0002-4203-9320}\inst{\ref{aff3}}
\and B.~Y.~Irureta-Goyena\orcid{0009-0004-5327-8767}\inst{\ref{aff16}}
\and B.~Simmons\orcid{0000-0001-5882-3323}\inst{\ref{aff17}}
\and J.~Reerink\orcid{0000-0001-6558-5183}\inst{\ref{aff2}}
\and H.~Dom\'inguez~S\'anchez\orcid{0000-0002-9013-1316}\inst{\ref{aff18}}
\and J.~H.~Knapen\orcid{0000-0003-1643-0024}\inst{\ref{aff19},\ref{aff20}}
\and F.~Shankar\orcid{0000-0001-8973-5051}\inst{\ref{aff21}}
\and E.~Duran-Camacho\orcid{0000-0002-3153-0536}\inst{\ref{aff19},\ref{aff20}}
\and J.~M.~P\'erez-Mart\'inez\orcid{0000-0002-5963-6850}\inst{\ref{aff19},\ref{aff20}}
\and H.~Dannerbauer\orcid{0000-0001-7147-3575}\inst{\ref{aff19},\ref{aff20}}
\and M.~Walmsley\orcid{0000-0002-6408-4181}\inst{\ref{aff22},\ref{aff23}}
\and N.~Auricchio\orcid{0000-0003-4444-8651}\inst{\ref{aff24}}
\and C.~Baccigalupi\orcid{0000-0002-8211-1630}\inst{\ref{aff10},\ref{aff9},\ref{aff25},\ref{aff26}}
\and M.~Baldi\orcid{0000-0003-4145-1943}\inst{\ref{aff27},\ref{aff24},\ref{aff28}}
\and S.~Bardelli\orcid{0000-0002-8900-0298}\inst{\ref{aff24}}
\and P.~Battaglia\orcid{0000-0002-7337-5909}\inst{\ref{aff24}}
\and A.~Biviano\orcid{0000-0002-0857-0732}\inst{\ref{aff9},\ref{aff10}}
\and M.~Bolzonella\orcid{0000-0003-3278-4607}\inst{\ref{aff24}}
\and A.~Bonchi\orcid{0000-0002-2667-5482}\inst{\ref{aff29}}
\and E.~Branchini\orcid{0000-0002-0808-6908}\inst{\ref{aff30},\ref{aff31},\ref{aff7}}
\and M.~Brescia\orcid{0000-0001-9506-5680}\inst{\ref{aff32},\ref{aff33}}
\and J.~Brinchmann\orcid{0000-0003-4359-8797}\inst{\ref{aff34},\ref{aff35},\ref{aff36}}
\and S.~Camera\orcid{0000-0003-3399-3574}\inst{\ref{aff37},\ref{aff38},\ref{aff39}}
\and G.~Ca\~nas-Herrera\orcid{0000-0003-2796-2149}\inst{\ref{aff4}}
\and V.~Capobianco\orcid{0000-0002-3309-7692}\inst{\ref{aff39}}
\and C.~Carbone\orcid{0000-0003-0125-3563}\inst{\ref{aff40}}
\and J.~Carretero\orcid{0000-0002-3130-0204}\inst{\ref{aff41},\ref{aff42}}
\and M.~Castellano\orcid{0000-0001-9875-8263}\inst{\ref{aff43}}
\and G.~Castignani\orcid{0000-0001-6831-0687}\inst{\ref{aff24}}
\and S.~Cavuoti\orcid{0000-0002-3787-4196}\inst{\ref{aff33},\ref{aff44}}
\and A.~Cimatti\inst{\ref{aff45}}
\and C.~Colodro-Conde\inst{\ref{aff19}}
\and G.~Congedo\orcid{0000-0003-2508-0046}\inst{\ref{aff46}}
\and C.~J.~Conselice\orcid{0000-0003-1949-7638}\inst{\ref{aff23}}
\and L.~Conversi\orcid{0000-0002-6710-8476}\inst{\ref{aff47},\ref{aff2}}
\and Y.~Copin\orcid{0000-0002-5317-7518}\inst{\ref{aff48}}
\and A.~Costille\inst{\ref{aff49}}
\and F.~Courbin\orcid{0000-0003-0758-6510}\inst{\ref{aff50},\ref{aff51},\ref{aff52}}
\and H.~M.~Courtois\orcid{0000-0003-0509-1776}\inst{\ref{aff53}}
\and M.~Cropper\orcid{0000-0003-4571-9468}\inst{\ref{aff54}}
\and H.~Degaudenzi\orcid{0000-0002-5887-6799}\inst{\ref{aff55}}
\and G.~De~Lucia\orcid{0000-0002-6220-9104}\inst{\ref{aff9}}
\and H.~Dole\orcid{0000-0002-9767-3839}\inst{\ref{aff14}}
\and F.~Dubath\orcid{0000-0002-6533-2810}\inst{\ref{aff55}}
\and X.~Dupac\inst{\ref{aff2}}
\and S.~Dusini\orcid{0000-0002-1128-0664}\inst{\ref{aff56}}
\and A.~Ealet\orcid{0000-0003-3070-014X}\inst{\ref{aff48}}
\and M.~Fabricius\orcid{0000-0002-7025-6058}\inst{\ref{aff57},\ref{aff58}}
\and M.~Farina\orcid{0000-0002-3089-7846}\inst{\ref{aff59}}
\and R.~Farinelli\inst{\ref{aff24}}
\and S.~Ferriol\inst{\ref{aff48}}
\and P.~Fosalba\orcid{0000-0002-1510-5214}\inst{\ref{aff60},\ref{aff61}}
\and S.~Fotopoulou\orcid{0000-0002-9686-254X}\inst{\ref{aff62}}
\and M.~Frailis\orcid{0000-0002-7400-2135}\inst{\ref{aff9}}
\and E.~Franceschi\orcid{0000-0002-0585-6591}\inst{\ref{aff24}}
\and M.~Fumana\orcid{0000-0001-6787-5950}\inst{\ref{aff40}}
\and S.~Galeotta\orcid{0000-0002-3748-5115}\inst{\ref{aff9}}
\and K.~George\orcid{0000-0002-1734-8455}\inst{\ref{aff63}}
\and B.~Gillis\orcid{0000-0002-4478-1270}\inst{\ref{aff46}}
\and C.~Giocoli\orcid{0000-0002-9590-7961}\inst{\ref{aff24},\ref{aff28}}
\and P.~G\'omez-Alvarez\orcid{0000-0002-8594-5358}\inst{\ref{aff64},\ref{aff2}}
\and J.~Gracia-Carpio\orcid{0000-0003-4689-3134}\inst{\ref{aff57}}
\and A.~Grazian\orcid{0000-0002-5688-0663}\inst{\ref{aff65}}
\and F.~Grupp\inst{\ref{aff57},\ref{aff58}}
\and S.~Gwyn\orcid{0000-0001-8221-8406}\inst{\ref{aff66}}
\and W.~Holmes\orcid{0009-0007-8554-4646}\inst{\ref{aff67}}
\and I.~M.~Hook\orcid{0000-0002-2960-978X}\inst{\ref{aff17}}
\and F.~Hormuth\inst{\ref{aff68}}
\and A.~Hornstrup\orcid{0000-0002-3363-0936}\inst{\ref{aff69},\ref{aff70}}
\and M.~Huertas-Company\orcid{0000-0002-1416-8483}\inst{\ref{aff19},\ref{aff71},\ref{aff72}}
\and K.~Jahnke\orcid{0000-0003-3804-2137}\inst{\ref{aff73}}
\and M.~Jhabvala\inst{\ref{aff74}}
\and B.~Joachimi\orcid{0000-0001-7494-1303}\inst{\ref{aff75}}
\and S.~Kermiche\orcid{0000-0002-0302-5735}\inst{\ref{aff76}}
\and A.~Kiessling\orcid{0000-0002-2590-1273}\inst{\ref{aff67}}
\and B.~Kubik\orcid{0009-0006-5823-4880}\inst{\ref{aff48}}
\and M.~K\"ummel\orcid{0000-0003-2791-2117}\inst{\ref{aff58}}
\and M.~Kunz\orcid{0000-0002-3052-7394}\inst{\ref{aff77}}
\and H.~Kurki-Suonio\orcid{0000-0002-4618-3063}\inst{\ref{aff78},\ref{aff79}}
\and A.~M.~C.~Le~Brun\orcid{0000-0002-0936-4594}\inst{\ref{aff80}}
\and S.~Ligori\orcid{0000-0003-4172-4606}\inst{\ref{aff39}}
\and P.~B.~Lilje\orcid{0000-0003-4324-7794}\inst{\ref{aff81}}
\and V.~Lindholm\orcid{0000-0003-2317-5471}\inst{\ref{aff78},\ref{aff79}}
\and I.~Lloro\orcid{0000-0001-5966-1434}\inst{\ref{aff82}}
\and G.~Mainetti\orcid{0000-0003-2384-2377}\inst{\ref{aff83}}
\and O.~Mansutti\orcid{0000-0001-5758-4658}\inst{\ref{aff9}}
\and O.~Marggraf\orcid{0000-0001-7242-3852}\inst{\ref{aff84}}
\and M.~Martinelli\orcid{0000-0002-6943-7732}\inst{\ref{aff43},\ref{aff85}}
\and N.~Martinet\orcid{0000-0003-2786-7790}\inst{\ref{aff49}}
\and F.~Marulli\orcid{0000-0002-8850-0303}\inst{\ref{aff86},\ref{aff24},\ref{aff28}}
\and R.~J.~Massey\orcid{0000-0002-6085-3780}\inst{\ref{aff87}}
\and S.~Maurogordato\inst{\ref{aff88}}
\and E.~Medinaceli\orcid{0000-0002-4040-7783}\inst{\ref{aff24}}
\and M.~Melchior\inst{\ref{aff89}}
\and M.~Meneghetti\orcid{0000-0003-1225-7084}\inst{\ref{aff24},\ref{aff28}}
\and E.~Merlin\orcid{0000-0001-6870-8900}\inst{\ref{aff43}}
\and G.~Meylan\inst{\ref{aff16}}
\and A.~Mora\orcid{0000-0002-1922-8529}\inst{\ref{aff90}}
\and M.~Moresco\orcid{0000-0002-7616-7136}\inst{\ref{aff86},\ref{aff24}}
\and L.~Moscardini\orcid{0000-0002-3473-6716}\inst{\ref{aff86},\ref{aff24},\ref{aff28}}
\and E.~Munari\orcid{0000-0002-1751-5946}\inst{\ref{aff9},\ref{aff10}}
\and C.~Neissner\orcid{0000-0001-8524-4968}\inst{\ref{aff91},\ref{aff42}}
\and S.-M.~Niemi\orcid{0009-0005-0247-0086}\inst{\ref{aff3}}
\and J.~W.~Nightingale\orcid{0000-0002-8987-7401}\inst{\ref{aff92}}
\and C.~Padilla\orcid{0000-0001-7951-0166}\inst{\ref{aff91}}
\and S.~Paltani\orcid{0000-0002-8108-9179}\inst{\ref{aff55}}
\and F.~Pasian\orcid{0000-0002-4869-3227}\inst{\ref{aff9}}
\and K.~Pedersen\inst{\ref{aff93}}
\and W.~J.~Percival\orcid{0000-0002-0644-5727}\inst{\ref{aff94},\ref{aff95},\ref{aff96}}
\and A.~Pezzotta\orcid{0000-0003-0726-2268}\inst{\ref{aff7}}
\and S.~Pires\orcid{0000-0002-0249-2104}\inst{\ref{aff97}}
\and G.~Polenta\orcid{0000-0003-4067-9196}\inst{\ref{aff29}}
\and M.~Poncet\inst{\ref{aff98}}
\and L.~A.~Popa\inst{\ref{aff99}}
\and L.~Pozzetti\orcid{0000-0001-7085-0412}\inst{\ref{aff24}}
\and F.~Raison\orcid{0000-0002-7819-6918}\inst{\ref{aff57}}
\and A.~Renzi\orcid{0000-0001-9856-1970}\inst{\ref{aff100},\ref{aff56},\ref{aff24}}
\and J.~Rhodes\orcid{0000-0002-4485-8549}\inst{\ref{aff67}}
\and G.~Riccio\inst{\ref{aff33}}
\and E.~Romelli\orcid{0000-0003-3069-9222}\inst{\ref{aff9}}
\and M.~Roncarelli\orcid{0000-0001-9587-7822}\inst{\ref{aff24}}
\and B.~Rusholme\orcid{0000-0001-7648-4142}\inst{\ref{aff101}}
\and R.~Saglia\orcid{0000-0003-0378-7032}\inst{\ref{aff58},\ref{aff57}}
\and Z.~Sakr\orcid{0000-0002-4823-3757}\inst{\ref{aff102},\ref{aff103},\ref{aff104}}
\and A.~G.~S\'anchez\orcid{0000-0003-1198-831X}\inst{\ref{aff57}}
\and D.~Sapone\orcid{0000-0001-7089-4503}\inst{\ref{aff105}}
\and B.~Sartoris\orcid{0000-0003-1337-5269}\inst{\ref{aff58},\ref{aff9}}
\and P.~Schneider\orcid{0000-0001-8561-2679}\inst{\ref{aff84}}
\and T.~Schrabback\orcid{0000-0002-6987-7834}\inst{\ref{aff106}}
\and M.~Scodeggio\inst{\ref{aff40}}
\and A.~Secroun\orcid{0000-0003-0505-3710}\inst{\ref{aff76}}
\and E.~Sihvola\orcid{0000-0003-1804-7715}\inst{\ref{aff107}}
\and P.~Simon\inst{\ref{aff84}}
\and C.~Sirignano\orcid{0000-0002-0995-7146}\inst{\ref{aff100},\ref{aff56}}
\and G.~Sirri\orcid{0000-0003-2626-2853}\inst{\ref{aff28}}
\and P.~Tallada-Cresp\'{i}\orcid{0000-0002-1336-8328}\inst{\ref{aff41},\ref{aff42}}
\and A.~N.~Taylor\inst{\ref{aff46}}
\and H.~I.~Teplitz\orcid{0000-0002-7064-5424}\inst{\ref{aff108}}
\and I.~Tereno\orcid{0000-0002-4537-6218}\inst{\ref{aff109},\ref{aff110}}
\and N.~Tessore\orcid{0000-0002-9696-7931}\inst{\ref{aff54}}
\and S.~Toft\orcid{0000-0003-3631-7176}\inst{\ref{aff111},\ref{aff112}}
\and R.~Toledo-Moreo\orcid{0000-0002-2997-4859}\inst{\ref{aff113}}
\and F.~Torradeflot\orcid{0000-0003-1160-1517}\inst{\ref{aff42},\ref{aff41}}
\and I.~Tutusaus\orcid{0000-0002-3199-0399}\inst{\ref{aff61},\ref{aff60},\ref{aff103}}
\and J.~Valiviita\orcid{0000-0001-6225-3693}\inst{\ref{aff78},\ref{aff79}}
\and T.~Vassallo\orcid{0000-0001-6512-6358}\inst{\ref{aff9},\ref{aff63}}
\and G.~Verdoes~Kleijn\orcid{0000-0001-5803-2580}\inst{\ref{aff114}}
\and A.~Veropalumbo\orcid{0000-0003-2387-1194}\inst{\ref{aff7},\ref{aff31},\ref{aff30}}
\and Y.~Wang\orcid{0000-0002-4749-2984}\inst{\ref{aff101}}
\and J.~Weller\orcid{0000-0002-8282-2010}\inst{\ref{aff58},\ref{aff57}}
\and A.~Zacchei\orcid{0000-0003-0396-1192}\inst{\ref{aff9},\ref{aff10}}
\and G.~Zamorani\orcid{0000-0002-2318-301X}\inst{\ref{aff24}}
\and F.~M.~Zerbi\orcid{0000-0002-9996-973X}\inst{\ref{aff7}}
\and E.~Zucca\orcid{0000-0002-5845-8132}\inst{\ref{aff24}}
\and J.~Macias-Perez\orcid{0000-0002-5385-2763}\inst{\ref{aff115}}
\and M.~Sereno\orcid{0000-0003-0302-0325}\inst{\ref{aff24},\ref{aff28}}}

\institute{Centro de Astrobiolog\'ia (CAB), CSIC-INTA, ESAC Campus, Camino Bajo del Castillo s/n, 28692 Villanueva de la Ca\~nada, Madrid, Spain\label{aff1}
\and
ESAC/ESA, Camino Bajo del Castillo, s/n., Urb. Villafranca del Castillo, 28692 Villanueva de la Ca\~nada, Madrid, Spain\label{aff2}
\and
European Space Agency/ESTEC, Keplerlaan 1, 2201 AZ Noordwijk, The Netherlands\label{aff3}
\and
Leiden Observatory, Leiden University, Einsteinweg 55, 2333 CC Leiden, The Netherlands\label{aff4}
\and
Universiteit Gent, Department of Physics and Astronomy, Proeftuinstraat 86 N3, 9000 Ghent, Belgium
\label{aff5}
\and
Dipartimento di Fisica ``G. Occhialini", Universit\`a degli Studi di Milano Bicocca, Piazza della Scienza 3, 20126 Milano, Italy\label{aff6}
\and
INAF-Osservatorio Astronomico di Brera, Via Brera 28, 20122 Milano, Italy\label{aff7}
\and
Dunlap Institute for Astronomy \& Astrophysics, University of Toronto, 50 St. George Street, Toronto, ON M5S 3H4, Canada\label{aff8}
\and
INAF-Osservatorio Astronomico di Trieste, Via G. B. Tiepolo 11, 34143 Trieste, Italy\label{aff9}
\and
IFPU, Institute for Fundamental Physics of the Universe, via Beirut 2, 34151 Trieste, Italy\label{aff10}
\and
Dipartimento di Fisica - Sezione di Astronomia, Universit\`a di Trieste, Via Tiepolo 11, 34131 Trieste, Italy\label{aff11}
\and
Department of Physics, University of Oxford, Keble Road, Oxford OX1 3RH, UK\label{aff12}
\and
Univ. Lille, CNRS, Centrale Lille, UMR 9189 CRIStAL, 59000 Lille, France\label{aff13}
\and
Universit\'e Paris-Saclay, CNRS, Institut d'astrophysique spatiale, 91405, Orsay, France\label{aff14}
\and
Institut d'Astrophysique de Paris, 98bis Boulevard Arago, 75014, Paris, France\label{aff15}
\and
Institute of Physics, Laboratory of Astrophysics, Ecole Polytechnique F\'ed\'erale de Lausanne (EPFL), Observatoire de Sauverny, 1290 Versoix, Switzerland\label{aff16}
\and
Department of Physics, Lancaster University, Lancaster, LA1 4YB, UK\label{aff17}
\and
Instituto de F\'isica de Cantabria, Edificio Juan Jord\'a, Avenida de los Castros, 39005 Santander, Spain\label{aff18}
\and
Instituto de Astrof\'{\i}sica de Canarias, E-38205 La Laguna, Tenerife, Spain\label{aff19}
\and
Universidad de La Laguna, Dpto. Astrof\'\i sica, E-38206 La Laguna, Tenerife, Spain\label{aff20}
\and
School of Physics \& Astronomy, University of Southampton, Highfield Campus, Southampton SO17 1BJ, UK\label{aff21}
\and
David A. Dunlap Department of Astronomy \& Astrophysics, University of Toronto, 50 St George Street, Toronto, Ontario M5S 3H4, Canada\label{aff22}
\and
Jodrell Bank Centre for Astrophysics, Department of Physics and Astronomy, University of Manchester, Oxford Road, Manchester M13 9PL, UK\label{aff23}
\and
INAF-Osservatorio di Astrofisica e Scienza dello Spazio di Bologna, Via Piero Gobetti 93/3, 40129 Bologna, Italy\label{aff24}
\and
INFN, Sezione di Trieste, Via Valerio 2, 34127 Trieste TS, Italy\label{aff25}
\and
SISSA, International School for Advanced Studies, Via Bonomea 265, 34136 Trieste TS, Italy\label{aff26}
\and
Dipartimento di Fisica e Astronomia, Universit\`a di Bologna, Via Gobetti 93/2, 40129 Bologna, Italy\label{aff27}
\and
INFN-Sezione di Bologna, Viale Berti Pichat 6/2, 40127 Bologna, Italy\label{aff28}
\and
Space Science Data Center, Italian Space Agency, via del Politecnico snc, 00133 Roma, Italy\label{aff29}
\and
Dipartimento di Fisica, Universit\`a di Genova, Via Dodecaneso 33, 16146, Genova, Italy\label{aff30}
\and
INFN-Sezione di Genova, Via Dodecaneso 33, 16146, Genova, Italy\label{aff31}
\and
Department of Physics "E. Pancini", University Federico II, Via Cinthia 6, 80126, Napoli, Italy\label{aff32}
\and
INAF-Osservatorio Astronomico di Capodimonte, Via Moiariello 16, 80131 Napoli, Italy\label{aff33}
\and
Instituto de Astrof\'isica e Ci\^encias do Espa\c{c}o, Universidade do Porto, CAUP, Rua das Estrelas, PT4150-762 Porto, Portugal\label{aff34}
\and
Faculdade de Ci\^encias da Universidade do Porto, Rua do Campo de Alegre, 4150-007 Porto, Portugal\label{aff35}
\and
European Southern Observatory, Karl-Schwarzschild-Str.~2, 85748 Garching, Germany\label{aff36}
\and
Dipartimento di Fisica, Universit\`a degli Studi di Torino, Via P. Giuria 1, 10125 Torino, Italy\label{aff37}
\and
INFN-Sezione di Torino, Via P. Giuria 1, 10125 Torino, Italy\label{aff38}
\and
INAF-Osservatorio Astrofisico di Torino, Via Osservatorio 20, 10025 Pino Torinese (TO), Italy\label{aff39}
\and
INAF-IASF Milano, Via Alfonso Corti 12, 20133 Milano, Italy\label{aff40}
\and
Centro de Investigaciones Energ\'eticas, Medioambientales y Tecnol\'ogicas (CIEMAT), Avenida Complutense 40, 28040 Madrid, Spain\label{aff41}
\and
Port d'Informaci\'{o} Cient\'{i}fica, Campus UAB, C. Albareda s/n, 08193 Bellaterra (Barcelona), Spain\label{aff42}
\and
INAF-Osservatorio Astronomico di Roma, Via Frascati 33, 00078 Monteporzio Catone, Italy\label{aff43}
\and
INFN -- Sezione di Napoli, Via Cinthia 6, 80126, Napoli, Italy\label{aff44}
\and
Dipartimento di Fisica e Astronomia "Augusto Righi" - Alma Mater Studiorum Universit\`a di Bologna, Viale Berti Pichat 6/2, 40127 Bologna, Italy\label{aff45}
\and
Institute for Astronomy, University of Edinburgh, Royal Observatory, Blackford Hill, Edinburgh EH9 3HJ, UK\label{aff46}
\and
European Space Agency/ESRIN, Largo Galileo Galilei 1, 00044 Frascati, Roma, Italy\label{aff47}
\and
Universit\'e Claude Bernard Lyon 1, CNRS/IN2P3, IP2I Lyon, UMR 5822, Villeurbanne, F-69100, France\label{aff48}
\and
Aix-Marseille Universit\'e, CNRS, CNES, LAM, Marseille, France\label{aff49}
\and
Institut de Ci\`{e}ncies del Cosmos (ICCUB), Universitat de Barcelona (IEEC-UB), Mart\'{i} i Franqu\`{e}s 1, 08028 Barcelona, Spain\label{aff50}
\and
Instituci\'o Catalana de Recerca i Estudis Avan\c{c}ats (ICREA), Passeig de Llu\'{\i}s Companys 23, 08010 Barcelona, Spain\label{aff51}
\and
Institut de Ciencies de l'Espai (IEEC-CSIC), Campus UAB, Carrer de Can Magrans, s/n Cerdanyola del Vall\'es, 08193 Barcelona, Spain\label{aff52}
\and
UCB Lyon 1, CNRS/IN2P3, IUF, IP2I Lyon, 4 rue Enrico Fermi, 69622 Villeurbanne, France\label{aff53}
\and
Mullard Space Science Laboratory, University College London, Holmbury St Mary, Dorking, Surrey RH5 6NT, UK\label{aff54}
\and
Department of Astronomy, University of Geneva, ch. d'Ecogia 16, 1290 Versoix, Switzerland\label{aff55}
\and
INFN-Padova, Via Marzolo 8, 35131 Padova, Italy\label{aff56}
\and
Max Planck Institute for Extraterrestrial Physics, Giessenbachstr. 1, 85748 Garching, Germany\label{aff57}
\and
Universit\"ats-Sternwarte M\"unchen, Fakult\"at f\"ur Physik, Ludwig-Maximilians-Universit\"at M\"unchen, Scheinerstr.~1, 81679 M\"unchen, Germany\label{aff58}
\and
INAF-Istituto di Astrofisica e Planetologia Spaziali, via del Fosso del Cavaliere, 100, 00100 Roma, Italy\label{aff59}
\and
Institut d'Estudis Espacials de Catalunya (IEEC),  Edifici RDIT, Campus UPC, 08860 Castelldefels, Barcelona, Spain\label{aff60}
\and
Institute of Space Sciences (ICE, CSIC), Campus UAB, Carrer de Can Magrans, s/n, 08193 Barcelona, Spain\label{aff61}
\and
School of Physics, HH Wills Physics Laboratory, University of Bristol, Tyndall Avenue, Bristol, BS8 1TL, UK\label{aff62}
\and
University Observatory, LMU Faculty of Physics, Scheinerstr.~1, 81679 Munich, Germany\label{aff63}
\and
FRACTAL S.L.N.E., calle Tulip\'an 2, Portal 13 1A, 28231, Las Rozas de Madrid, Spain\label{aff64}
\and
INAF-Osservatorio Astronomico di Padova, Via dell'Osservatorio 5, 35122 Padova, Italy\label{aff65}
\and
National Research Council, Herzberg Astronomy and Astrophysics Research Centre, 5071 W. Saanich Rd. Victoria, BC, V9E 2E7, Canada\label{aff66}
\and
Jet Propulsion Laboratory, California Institute of Technology, 4800 Oak Grove Drive, Pasadena, CA, 91109, USA\label{aff67}
\and
Felix Hormuth Engineering, Goethestr. 17, 69181 Leimen, Germany\label{aff68}
\and
Technical University of Denmark, Elektrovej 327, 2800 Kgs. Lyngby, Denmark\label{aff69}
\and
Cosmic Dawn Center (DAWN), Denmark\label{aff70}
\and
Universit\'e PSL, Observatoire de Paris, Sorbonne Universit\'e, CNRS, LERMA, 75014, Paris, France\label{aff71}
\and
Universit\'e Paris-Cit\'e, 5 Rue Thomas Mann, 75013, Paris, France\label{aff72}
\and
Max-Planck-Institut f\"ur Astronomie, K\"onigstuhl 17, 69117 Heidelberg, Germany\label{aff73}
\and
NASA Goddard Space Flight Center, Greenbelt, MD 20771, USA\label{aff74}
\and
Department of Physics and Astronomy, University College London, Gower Street, London WC1E 6BT, UK\label{aff75}
\and
Aix-Marseille Universit\'e, CNRS/IN2P3, CPPM, Marseille, France\label{aff76}
\and
Universit\'e de Gen\`eve, D\'epartement de Physique Th\'eorique and Centre for Astroparticle Physics, 24 quai Ernest-Ansermet, CH-1211 Gen\`eve 4, Switzerland\label{aff77}
\and
Department of Physics, P.O. Box 64, University of Helsinki, 00014 Helsinki, Finland\label{aff78}
\and
Helsinki Institute of Physics, Gustaf H{\"a}llstr{\"o}min katu 2, University of Helsinki, 00014 Helsinki, Finland\label{aff79}
\and
Laboratoire d'etude de l'Univers et des phenomenes eXtremes, Observatoire de Paris, Universit\'e PSL, Sorbonne Universit\'e, CNRS, 92190 Meudon, France\label{aff80}
\and
Institute of Theoretical Astrophysics, University of Oslo, P.O. Box 1029 Blindern, 0315 Oslo, Norway\label{aff81}
\and
SKAO, Jodrell Bank, Lower Withington, Macclesfield SK11 9FT, UK\label{aff82}
\and
Centre de Calcul de l'IN2P3/CNRS, 21 avenue Pierre de Coubertin 69627 Villeurbanne Cedex, France\label{aff83}
\and
Universit\"at Bonn, Argelander-Institut f\"ur Astronomie, Auf dem H\"ugel 71, 53121 Bonn, Germany\label{aff84}
\and
INFN-Sezione di Roma, Piazzale Aldo Moro, 2 - c/o Dipartimento di Fisica, Edificio G. Marconi, 00185 Roma, Italy\label{aff85}
\and
Dipartimento di Fisica e Astronomia "Augusto Righi" - Alma Mater Studiorum Universit\`a di Bologna, via Piero Gobetti 93/2, 40129 Bologna, Italy\label{aff86}
\and
Department of Physics, Institute for Computational Cosmology, Durham University, South Road, Durham, DH1 3LE, UK\label{aff87}
\and
Universit\'e C\^{o}te d'Azur, Observatoire de la C\^{o}te d'Azur, CNRS, Laboratoire Lagrange, Bd de l'Observatoire, CS 34229, 06304 Nice cedex 4, France\label{aff88}
\and
University of Applied Sciences and Arts of Northwestern Switzerland, School of Engineering, 5210 Windisch, Switzerland\label{aff89}
\and
Telespazio UK S.L. for European Space Agency (ESA), Camino bajo del Castillo, s/n, Urbanizacion Villafranca del Castillo, Villanueva de la Ca\~nada, 28692 Madrid, Spain\label{aff90}
\and
Institut de F\'{i}sica d'Altes Energies (IFAE), The Barcelona Institute of Science and Technology, Campus UAB, 08193 Bellaterra (Barcelona), Spain\label{aff91}
\and
School of Mathematics, Statistics and Physics, Newcastle University, Herschel Building, Newcastle-upon-Tyne, NE1 7RU, UK\label{aff92}
\and
DARK, Niels Bohr Institute, University of Copenhagen, Jagtvej 155, 2200 Copenhagen, Denmark\label{aff93}
\and
Waterloo Centre for Astrophysics, University of Waterloo, Waterloo, Ontario N2L 3G1, Canada\label{aff94}
\and
Department of Physics and Astronomy, University of Waterloo, Waterloo, Ontario N2L 3G1, Canada\label{aff95}
\and
Perimeter Institute for Theoretical Physics, Waterloo, Ontario N2L 2Y5, Canada\label{aff96}
\and
Universit\'e Paris-Saclay, Universit\'e Paris Cit\'e, CEA, CNRS, AIM, 91191, Gif-sur-Yvette, France\label{aff97}
\and
Centre National d'Etudes Spatiales -- Centre spatial de Toulouse, 18 avenue Edouard Belin, 31401 Toulouse Cedex 9, France\label{aff98}
\and
Institute of Space Science, Str. Atomistilor, nr. 409 M\u{a}gurele, Ilfov, 077125, Romania\label{aff99}
\and
Dipartimento di Fisica e Astronomia "G. Galilei", Universit\`a di Padova, Via Marzolo 8, 35131 Padova, Italy\label{aff100}
\and
Caltech/IPAC, 1200 E. California Blvd., Pasadena, CA 91125, USA\label{aff101}
\and
Instituto de F\'isica Te\'orica UAM-CSIC, Campus de Cantoblanco, 28049 Madrid, Spain\label{aff102}
\and
Institut de Recherche en Astrophysique et Plan\'etologie (IRAP), Universit\'e de Toulouse, CNRS, UPS, CNES, 14 Av. Edouard Belin, 31400 Toulouse, France\label{aff103}
\and
Universit\'e St Joseph; Faculty of Sciences, Beirut, Lebanon\label{aff104}
\and
Departamento de F\'isica, FCFM, Universidad de Chile, Blanco Encalada 2008, Santiago, Chile\label{aff105}
\and
Universit\"at Innsbruck, Institut f\"ur Astro- und Teilchenphysik, Technikerstr. 25/8, 6020 Innsbruck, Austria\label{aff106}
\and
Department of Physics and Helsinki Institute of Physics, Gustaf H\"allstr\"omin katu 2, University of Helsinki, 00014 Helsinki, Finland\label{aff107}
\and
Infrared Processing and Analysis Center, California Institute of Technology, Pasadena, CA 91125, USA\label{aff108}
\and
Departamento de F\'isica, Faculdade de Ci\^encias, Universidade de Lisboa, Edif\'icio C8, Campo Grande, PT1749-016 Lisboa, Portugal\label{aff109}
\and
Instituto de Astrof\'isica e Ci\^encias do Espa\c{c}o, Faculdade de Ci\^encias, Universidade de Lisboa, Tapada da Ajuda, 1349-018 Lisboa, Portugal\label{aff110}
\and
Cosmic Dawn Center (DAWN)\label{aff111}
\and
Niels Bohr Institute, University of Copenhagen, Jagtvej 128, 2200 Copenhagen, Denmark\label{aff112}
\and
Universidad Polit\'ecnica de Cartagena, Departamento de Electr\'onica y Tecnolog\'ia de Computadoras,  Plaza del Hospital 1, 30202 Cartagena, Spain\label{aff113}
\and
Kapteyn Astronomical Institute, University of Groningen, PO Box 800, 9700 AV Groningen, The Netherlands\label{aff114}
\and
Univ. Grenoble Alpes, CNRS, Grenoble INP, LPSC-IN2P3, 53, Avenue des Martyrs, 38000 Grenoble, France\label{aff115}}      

   \date{}

  \abstract{Galaxy clusters provide unique laboratories for studying environmental effects on galaxy evolution. The morphology--density ($T$--$\Sigma$) and morphology--cluster-centric radius ($T$--$R$) relations trace how galaxy morphology is influenced by the environment, but previous studies at intermediate redshifts have been limited in both sample size and radial coverage. We used the Euclid Quick Data Release 1 (Q1) visual morphology catalog to measure the $T$--$\Sigma$ and $T$--$R$ relations for smooth, featured-or-disk, and barred galaxies in known clusters at $0.2\leq z\leq 0.5$, extending this analysis to large cluster-centric distances ($3\,R_{500c}$) and studying their dependence on stellar mass. Using photometric redshifts and stellar masses provided by \Euclid, we identify 1754 cluster members within $1.5\,R_{500c}$, distributed across 71 clusters. We classify the identified galaxies as smooth, featured-or-disk, or barred using the predicted vote fractions provided by the \Zoobot deep learning foundation model in the Q1 visual morphology catalog. Our \Zoobot classification of smooth and featured-or-disk galaxies is similar to the separation between elliptical and disk galaxies. We confirm the $T$--$\Sigma$ relation in all the stellar mass ranges studied, with a stronger influence of the cluster environment on galaxy morphology in the densest parts of the cluster and closer to the cluster center. Beyond $1.5\,R_{500c}$, the morphological segregation weakens, with featured-or-disk galaxies overtaking smooth galaxies at the lowest densities, consistent with the growing contribution of field interlopers in the cluster outskirts. For barred galaxies, we find a tentative decline of the bar fraction toward lower densities that is most pronounced for the most massive galaxies. In conclusion, our results show that the cluster environment drives a progressive transformation of galaxy morphology, with the loss of disk structure becoming more pronounced toward the cluster core, where environmental processes act most efficiently. These findings, enabled by the unique characteristics of the Q1 visual morphology catalog, advance our physical understanding of how galaxies evolve in dense environments at intermediate redshifts.}


   \maketitle

\section{Introduction}\label{sec:introduction}

The systematic classification of galaxies by their visual appearance has been a cornerstone of extragalactic astronomy since the pioneering work presented by \citet{hubble_1926}. The identification and analysis of morphological features such as bars, spiral arms, or bulge strength is fundamental to understanding the physical processes that drive galaxy formation and evolution over cosmic time \citep{de_vac_1959,kormendy_2010,buta_2015,huertas_2016,huertas_2024,tobias_2025}. In the modern era, large-scale citizen science projects such as Galaxy Zoo \citep{lintott_2008,masters_2019} have provided catalogs of hundreds of thousands of visually classified galaxies \citep{willett_2013,walmsley2022,gz_evo}, enabling a revolution in the study of visual morphological features \citep{simmons_2017,dominguez_2018,tobias_2021,shimakawa_2024,tobias_2025}. These projects are combined with state-of-the-art deep learning approaches, such as the \texttt{Zoobot} foundation model \citep{zoobot}, to reproduce and extend human classifications to the unprecedented scales of next-generation surveys \citep{walmsley2023,walmsley2024}.

Galaxy clusters represent unique laboratories for studying environmental effects on the evolution and morphology of galaxies. In these high-density environments, galaxies experience several physical processes, such as ram-pressure stripping \citep{gunn_1972,quilis_2000,boselli_2022}, tidal stripping \citep{merritt_1983,Natarajan_2009,pfeffer_2014}, galaxy harassment \citep{moore_1996,moore_1998,bialas_2015}, or strangulation \citep{larson_1980,tonnesen_2007,peng_2015}, which can alter their internal properties, such as star formation or morphology. Since it was first quantified by \citet{dressler_1980}, several studies have examined the relation between the local projected density of galaxies and the fraction of galaxies of different morphological types \citep[known as the $T$--$\Sigma$ relation;][]{goto_2003,postman_2005,bamford_2009,fasano_2015,mei_2023,vulcani2023}. Some of these have also focused on the relation between morphology and cluster-centric distance ($T$--$R$ relation)\footnote{Here, morphology is denoted as $T$ since it is related to the ``T-type'' numerical scheme of galaxy classification \citep{devac1991,buta_2015}. Although we use visual morphology classifications in this work, we still refer to the relations as $T$--$\Sigma$ and $T$--$R$.}. Although these two environmental indicators are correlated, they trace different aspects of the cluster environment and the physical processes taking place in it, since, for example, the effects of local overdensities within the cluster are not visible in the $T$--$R$ relation. Before entering the inner regions of the cluster environment, infalling galaxies could experience significant structural transformations in groups or filaments, a process known as ``preprocessing'' \citep{taranu_2014,sarron_2019}. Overall, these studies show that early-type galaxies (ETGs; ellipticals and S0s) are preferentially found in the densest regions and cluster cores, while the fraction of late-type disk galaxies (LTGs) increases toward lower-density regions and the cluster outskirts. 

The \Euclid space mission \citep{mellier_2025} is equipped with two instruments that uniquely combine high spatial resolution, sensitivity, and a large survey area, which will enable the study of the detailed visual morphology of galaxies on an unprecedented scale. The visible camera \citep[VIS; ][]{VIS} is a large optical-band imager with a single broadband filter, \IE, covering 550 to 920\,nm and achieving a 10$\sigma$ point-source sensitivity at 24.5\,mag. The Near-Infrared Spectrometer and Photometer \citep[NISP; ][]{NISP} provides near-infrared ($\simeq$900--2000 nm) photometry in the \YE, \JE, and \HE bands\footnote{Tabular versions of the individual filters are available at \url{https://doi.org/10.5270/esa-kx8w57c}} \citep{Schirmer-EP18}, each of which reaches a 5$\sigma$ point-source sensitivity at 24\,mag. It moreover provides low-resolution slitless spectroscopy with $R\ge450$. Covering 63\,deg$^{2}$, the visual morphology catalog \citep{zoobot_ec} of the \citet[][\citetalias{Q1}]{Q1}, the first \Euclid public data release, has already increased the number of galaxies with visual features at $0.3<z<0.7$ by an order of magnitude relative to similar Sloan Digital Sky Survey \citep{willett_2013} and the \textit{Hubble} Space Telescope catalogs \citep[HST; ][]{willett_2017}. The final Euclid Wide Survey \citep[EWS;][]{EWS}, which is expected to cover around 14\,000\,deg$^2$, will provide detailed visual morphology measurements for around $10^8$ galaxies up to $z\sim1.5$ \citep{bretonniere2023,aussel_2024}. The configuration and characteristics of the \Euclid mission offer two key strengths for our study. To date, it is the only survey that enables a statistically significant study of the $T$--$\Sigma$ relation in a sample of intermediate redshift clusters. Thanks to its large sky coverage, it also permits extending the study of the $T$--$R$ relation to large cluster-centric distances, mapping the cluster outskirts, where infalling and backsplash galaxies overlap with field interlopers, a major limitation of previous work.

This work is part of a series of papers dedicated to the study of visual galaxy morphology in large samples of galaxies in clusters. \citet{oryan_2026} leverages the capabilities offered by the ESA Datalabs platform\footnote{\url{https://datalabs.esa.int}}~\citep{Navarro2024_datalabs} to create a catalog of visual galaxy morphologies using the \texttt{Zoobot} model and source cutouts from the HST Legacy Archive. Subsequently, they use this catalog to recover the $T$--$\Sigma$ relation for $0.2\leq z\leq 0.6$ up to $R_{500c}$, limited by HST's field of view. Here, we complement this work by extending the study up to $3\,R_{500c}$ and to different stellar mass ranges, performing a comprehensive analysis of the change in the bar fraction with environment and stellar mass at intermediate redshifts. For this, we used a survey-oriented approach that significantly streamlines the process, without the need for ancillary  information.

In this paper, we use the Q1 visual morphology catalog to study the effects of cluster environment on galaxy morphology at $0.2\leq z\leq 0.5$, measuring the $T$--$\Sigma$ and $T$--$R$ relations in a sample of 1754 galaxies distributed across 71 clusters drawn from existing catalogs in the literature and extending these relations to the cluster outskirts. In Sect.~\ref{sec:data}, we describe the data used in this work. We present the methodology used to select our sample of galaxies and our classification schema in Sect.~\ref{sec:method}. In Sect.~\ref{sec:morph_results}, we discuss the results obtained for the relation between the detailed visual morphology and environment in clusters. Finally, we summarize the main conclusions of this work in Sect. \ref{sec:conclusions}. We adopt a flat $\Lambda$ cold dark matter cosmology following the \textit{Planck} mission \citep{planck_2018}, with $H_0=67.7$\,km\,s$^{-1}$\,Mpc$^{-1}$, $\Omega_\mathrm{m}=0.311$, and $\Omega_\Lambda=0.689$.

\section{Data}\label{sec:data}

\subsection{Euclid Quick Data Release 1}\label{sec:euclid_Q1}

This work is based on the Q1, with a sky coverage distributed across the three Euclid Deep Fields (EDFs): EDF-North (22.9 deg$^2$), EDF-South (28.1 deg$^2$), and EDF-Fornax (12.1 deg$^2$). With a high spatial resolution characterized by a full width at half maximum of \ang{;;.16} for the VIS Q1 point spread function \citep{mccracken2025}, the Q1 observations provide a preview of the typical single-visit depth of the EWS, which is expected to cover up to 14\,000 deg$^2$. We refer the reader to \citet{Q1_overview} for a comprehensive overview of the Q1 data processing and observations.

\begin{figure*}
    \centering
    	\includegraphics[width=.95\linewidth]{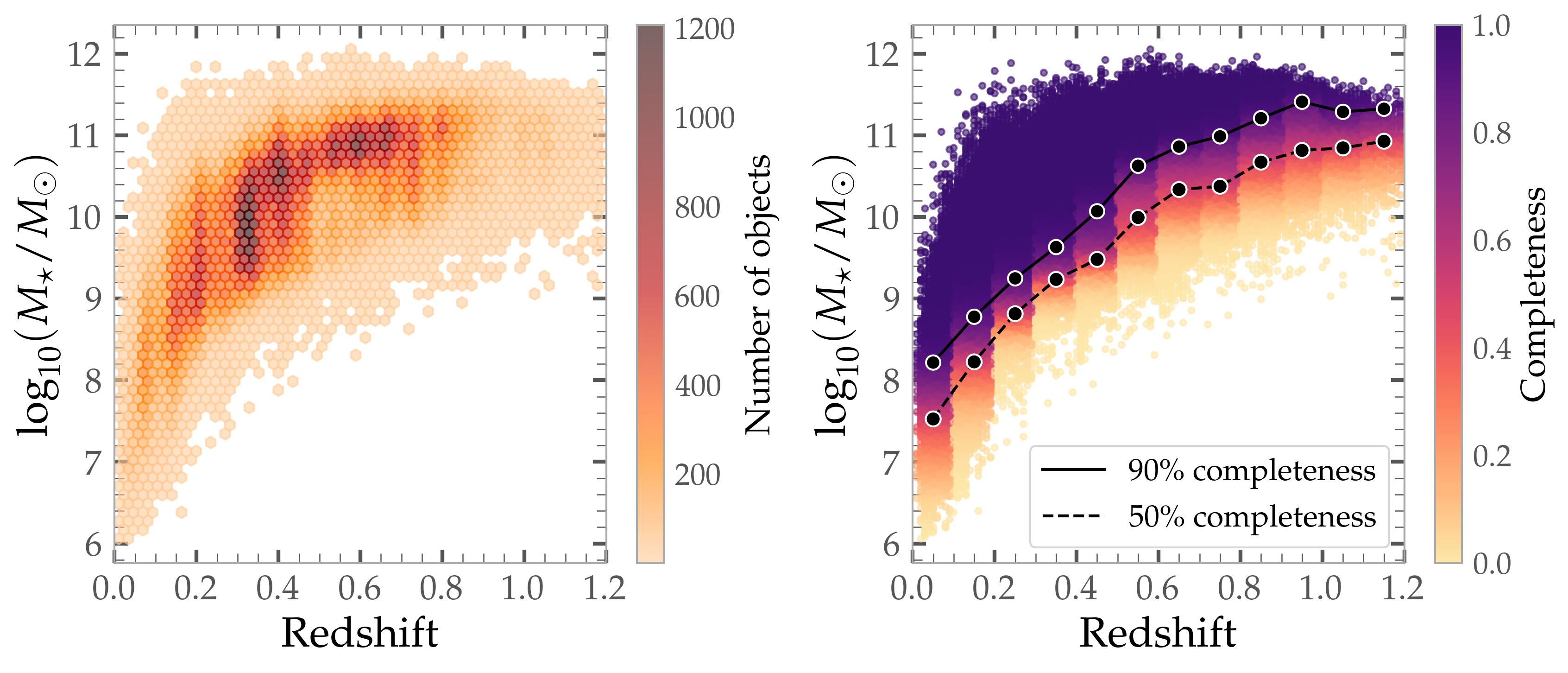}

    \caption{Stellar mass vs. photometric redshift diagram for galaxies included in our initial sample. \textit{Left panel:} Darker colours indicate higher number of galaxies inside the hexagonal bins. \textit{Right panel:} Dots color-coded by the completeness level estimated using the method developed by \citet{pozzetti2010}. The solid and dash-dotted lines correspond to the completeness limits at the 90\% and 50\% confidence levels, respectively.}
    \label{fig:comp}
\end{figure*}

The MERge processing function \citep[MER PF;][]{Q1-TP004} provides source catalogs with multiwavelength photometric data and a variety of morphological measurements, including parametric Sérsic fits, nonparametric morphologies, and deep learning-based visual morphologies obtained using the \texttt{Zoobot} model. The \texttt{Zoobot} morphological measurements included in the Q1 MER catalog are provided for galaxies that meet one of the following criteria:
\bi
    \itemindent=-15pt
    \item[]$\texttt{segmentation\_area}>1200$\,pixels
    \item[] \texttt{OR}
    \item[] $\texttt{segmentation\_area}>200  $\,pixels and $\IE<20.5$,
\ei
\noindent where \texttt{segmentation\_area} is defined as the total number of pixels within the segmentation source mask from \texttt{SourceXtractor++}, as calculated by \citet{Q1-TP004}. Besides these measurements, \citet{zoobot_ec} created the Q1 dynamic visual morphology catalog\footnote{\url{https://doi.org/10.5281/zenodo.15002907}}, in which they reduced the first criterion to $\texttt{segmentation\_area}>700$\,pixels and provided robust visual morphology measurements for a total of 380\,111 MER galaxies. This dynamic catalog, the one used in this work, was developed using an approach that combines deep transfer learning and citizen science. Within a dedicated campaign, 9976 Galaxy Zoo volunteers contributed 2.9 million classifications of 114\,000 \Euclid galaxies, answering a tree of questions (e.g., ``Is this galaxy smooth or featured?'', ``Does it have spiral arms?'') in which each subsequent question depends on the previous answer\footnote{The full tree schema is provided in the \texttt{zoobot.shared.schemas.euclid\_schema} Python module.}. Volunteers were shown an $\IE/\YE$ composite RGB image, a standard \IE-only grayscale image, and an \IE grayscale image adjusted to highlight low-surface brightness features. Then, the volunteer annotations were used to fine-tune the \texttt{Zoobot} foundation model, designed to be quickly transferable for downstream tasks in new surveys. As a result, the catalog provides \texttt{Zoobot}-predicted vote fractions for each of the morphology questions included in the tree schema, representing the model predictions of how volunteers would classify each galaxy based on its visual morphology.

We complemented the morphological measurements with the photometric redshifts and stellar masses provided by the PHotometric redshift processing function \citep[PHZ PF;][]{ou-phz}, which computes photometric redshifts and galaxy physical properties using two different methods. The \texttt{Phosphoros} template-fitting package provides Bayesian posterior distributions of the photometric redshifts, together with the median and the 1$\sigma$ uncertainties. The Nearest-Neighbour Photometric Redshifts (\texttt{NNPZ}) machine learning method provides galaxy physical properties using the 30 nearest neighbors from a reference sample of spectral energy distributions, assuming a Kroupa initial mass function \citep{kroupa_2001}. In this work, we used the median redshifts (\texttt{phz\_median}) and redshift probability density functions provided by \texttt{Phosphoros} and the median of the posterior distributions provided by \texttt{NNPZ} for the galaxy stellar masses. 

To ensure the reliability of photometric redshifts and the galaxy physical properties, we followed an approach inspired by those presented by \citet{cleland_ec} and \citet{quilley_ec} and imposed the following selection criteria:

\bi
    \item \texttt{phz\_flags} = 0,
    \item \texttt{phys\_param\_flags} = 0,
    \item one of the following redshift criteria is fulfilled:
    \begingroup\raggedright
    \bi
        \item[*] $\left|\texttt{phz\_median} - \texttt{phz\_pp\_median\_z}\right| <\allowbreak 0.1\,(1+\texttt{phz\_median})$,
        \item[*] \texttt{phz\_median} is within the 1$\sigma$ interval of \texttt{phz\_pp\_median\_z}, or vice versa, with a 1$\sigma$ interval smaller than 0.2,
    \ei
    \endgroup
\ei

\noindent where \texttt{phz\_pp\_median\_z} is the photometric redshift estimate from \texttt{NNPZ}. 

After removing 2622 objects with a \Zoobot expected volunteer vote fraction of \texttt{smooth-or-featured\_problem\_fraction} $>0.5$, likely artifacts (Fig. \ref{fig:prob}), the resulting sample yields 228\,633 galaxies, with a median photometric redshift uncertainty of $\sigma_z/(1+z)=0.03$. Of these galaxies, 60\% satisfy both redshift criteria simultaneously, 39.5\% satisfy only the first condition, and 0.5\% were selected exclusively through the second. The first condition is therefore the primary driver of the selection, while the second acts as a small recovery mechanism for galaxies with consistent photometric redshift posteriors from \texttt{Phosphoros} and \texttt{NNPZ}. Figure~\ref{fig:comp} shows the stellar mass versus photometric redshift plane for the sample obtained with the selection criteria discussed in this section.

The selection cuts used to create the Q1 dynamic visual morphology catalog were chosen to ensure robust morphological measurements, but they have a significant impact on sample completeness, as the smallest and faintest sources (with $\texttt{segmentation\_area}\leq 700$\,pixels and $\IE\geq20.5$) are excluded. As illustrated by Fig.~\ref{fig:comp}, the low-mass population becomes progressively less represented with increasing redshift, which makes it essential to account for this effect to avoid biasing the scientific analysis toward the brightest and most massive galaxies. Using the method developed by \citet{pozzetti2010}, we computed the completeness level for each of the objects at its corresponding redshift and stellar mass in redshift bins of 0.1 and derived the completeness limits at the 90\% and 50\% confidence levels for each of the bins (right panel of Fig.~\ref{fig:comp}). Our analysis of the morphological trends presented in Sect.~\ref{sec:morph_results} takes these measurements into account and adopts narrow stellar mass bins to mitigate the impact of sample incompleteness on our results.

\subsection{Cluster compilation}\label{sec:clusters}

We focused on a cluster compilation similar to the one presented by \citet{gouin_ec}, including eROSITA \citep{erass}, MCXC-II \citep{mcxc2}, the redMaPPer DES SVA1 expanded catalog \citep{redmapper}, and WHL-SDSS \citep{whl_sdss}. After de-duplicating clusters that appeared in multiple catalogs, removing those located on the edge of the Euclid Q1 footprint, and discarding those without $R_{500c}$ or $R_{200c}$ values in the literature catalogs, which removes all the redMaPPer clusters, we ended up with a sample of 254 clusters. This sample includes 118, 106, and 30 clusters from WHL-SDSS, eROSITA, and MCXC-II, respectively. With redshifts between 0.03 and 0.95, 200 of these clusters rely on photometric redshifts, while spectroscopic redshift is provided for 54 of them. The values of $R_{500c}$ (or $R_{200c}$) for each cluster were taken directly from these literature catalogs. When only $R_{200c}$ was provided, we estimated $R_{500c}$ by assuming a ratio $R_{500c}/R_{200c}=0.7$ \citep{ettori2009,gouin_ec}.


\section{Methodology}\label{sec:method}

\subsection{Cluster membership}\label{sec:members}

\begin{figure}
    \centering
    	\includegraphics[width=\linewidth]{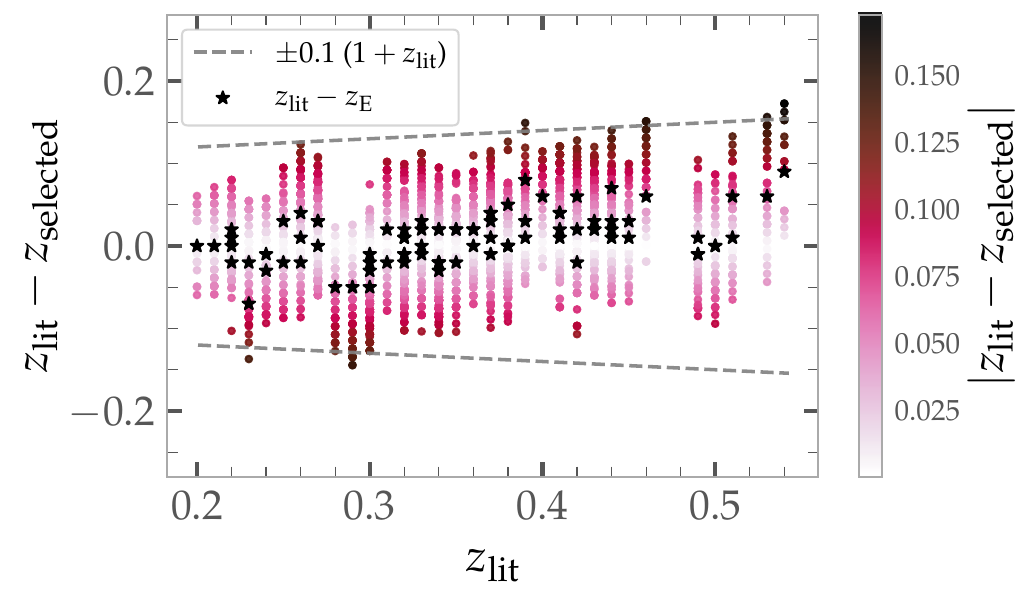}

    \caption{Comparison between the \Euclid photometric redshifts, $z_{\mathrm{selected}}$, of the galaxies satisfying the redshift membership criteria described in Sect. \ref{sec:members} and the cluster redshifts provided in the literature, $z_{\mathrm{lit}}$. The black stars represent the difference between $z_{\mathrm{lit}}$ and the computed \Euclid cluster redshift, $z_{\mathrm{E}}$.}
    \label{fig:z_cl_summary}
\end{figure}

\begin{table*}
 \caption{Steps followed to select the final sample of cluster members.}
 \label{tab:selection}
 \centering          
 \begin{tabular}{l c c c}
  \hline\hline
  \noalign{\smallskip}
  
  Sample & Selection & \# Galaxies & \# Clusters\\
  
  \noalign{\smallskip}
  \hline
  \noalign{\smallskip}

  Morphology catalog & \ldots & 380\,111 & \ldots \\

  \noalign{\smallskip}
  
  Initial sample & Selection criteria discussed in Sect. \ref{sec:data} & 228\,633 & \ldots\\

  \noalign{\smallskip}

  Positional match & $R\leq3\,R_{500c}$  & 54\,298 & 254 \\
  
  \noalign{\smallskip}

  Redshift cluster membership & Redshift selection criteria discussed in Sect. \ref{sec:members} &  6068 & 114 \\
  
  \noalign{\smallskip}

   Final sample of cluster members  & $0.2\leq z_{\mathrm{E}} \leq 0.5$ \& $R\leq1.5\,R_{500c}$ & 1754 & 71 \\  
  
  \noalign{\smallskip}  
  \hline
 \end{tabular}
\end{table*}

We first performed a positional match to obtain all objects in the initial sample that lie within $3\,R_{500c}$ (i.e., $\sim2\,R_{200c}$) of a cluster. For objects lying within $3\,R_{500c}$ of two or more clusters, we assigned the match to the cluster with the redshift closest to that of the object.

Then, we looked for a peak in the redshift space using a window of $\delta_z(1+z_{\mathrm{lit}})$ centered on $z_{\mathrm{lit}}$, where $z_{\mathrm{lit}}$ is the redshift of the cluster provided in the literature and $\delta_z$ is set to 0.05 and 0.1 for spectroscopic and photometric literature redshifts, respectively. Thus, we defined a \Euclid cluster redshift, $z_{\mathrm{E}}$, as the median of the distribution of \Euclid redshifts found within this window. The purpose of this step is to address both the scatter between \Euclid photometric redshifts and those provided in the literature \citep{bhargava_ec}, as well as contamination from other possible over-concentrations in the redshift space. We estimated membership probabilities, $P_{\mathrm{mem}}$, for each galaxy by integrating the photometric redshift probability density function within a selection window of $0.075(1+z_{\mathrm{E}})$ centered on $z_{\mathrm{E}}$, keeping only galaxies with $P_{\mathrm{mem}}>0.7$. This ensures that most of the redshift probability density of each selected galaxy lies within the selected cluster redshift window, reducing contamination from foreground and background sources scattered into the selection by their photometric redshift uncertainties. The comparison between the \Euclid photometric redshifts of the galaxies selected in this way and the cluster redshifts provided in the literature is illustrated in Fig.~\ref{fig:z_cl_summary}, showing how we used this approach to obtain a sample of galaxies from the \Euclid redshift space that is consistent with the values given in the literature for the cluster redshift. We discarded clusters with less than 20 objects within $3\,R_{500c}$ to ensure the reliability of the determination of $z_{\mathrm{E}}$.

\begin{figure}
    \centering
    	\includegraphics[width=.95\linewidth]{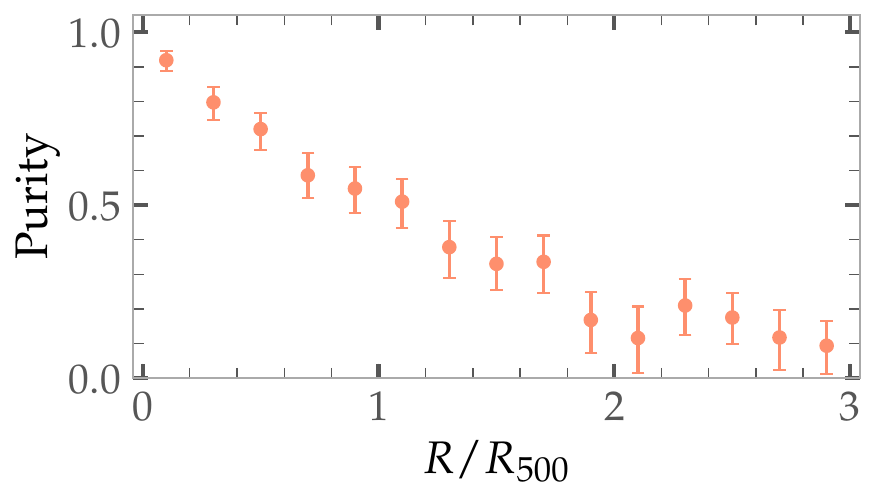}

    \caption{Purity as a function of cluster-centric distance, normalized to $R_{500c}$, for our membership selection. The error bars represent the 68\% percentiles from 1000 bootstrap samples.}
    \label{fig:purity}
\end{figure}

\begin{figure}
    \centering
    	\includegraphics[width=.95\linewidth]{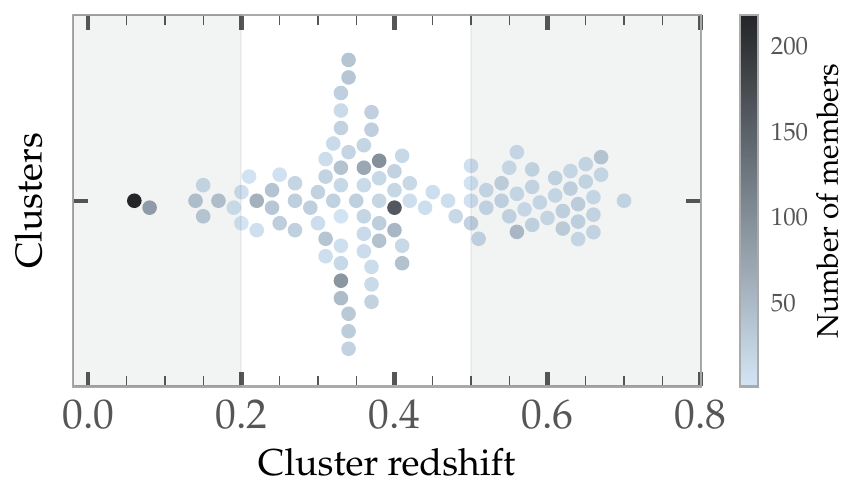}

    \caption{Swarm plot showing the distribution of the \Euclid cluster redshifts for the clusters with the galaxies selected in Sect. \ref{sec:members}. Arranged so they do not overlap and color-coded according to the galaxies assigned to each cluster, the dots represent individual clusters. The white area represents the clusters with $0.2\leq z\leq 0.5$ used for our study of galaxy morphology.}
    \label{fig:z_cl}
\end{figure}

Selecting cluster members using photometric redshifts inevitably allows some contamination from foreground and background sources. Consequently, the purity of the member sample declines toward the cluster outskirts, where the projected field density dominates over the cluster signal \citep{george_2011,castignani_benoist}. To quantify this for our sample, we estimated the purity of our selection as a function of cluster-centric distance. Following the notation adopted by \citet{george_2011} and \citet{castignani_benoist}, we defined the purity in a given radial bin as $p = 1 -N_{\mathrm{field}}/N_{\mathrm{mem}}$, where $N_{\mathrm{mem}}$ is the number of selected galaxies and $N_{\mathrm{field}}$ is the expected number of interlopers. We estimated $N_{\mathrm{field}}$ by repeating the exact selection described above on control fields: for each cluster, we placed apertures of the same angular size ($3\,R_{500c}$) on ten random pointings outside the cluster, each offset by 2.5 times the cluster radius, and applied the same redshift window and $P_{\mathrm{mem}}$ threshold, so that the galaxies that pass the selection represent the field contamination expected per radial bin. We averaged the field counts over the ten pointings for each cluster, and estimated the corresponding uncertainties using a cluster-level bootstrap, resampling the clusters with replacement and drawing one of the ten random pointings for each resampled cluster. We computed the purity profile for each of the 1000 bootstrap samples and adopted the 68\% percentiles as the uncertainty. Figure~\ref{fig:purity} shows that, as expected, the purity remains high in the cluster core and decreases with cluster-centric distance, dropping to $p\approx0.1$--$0.2$ beyond $2\,R_{500c}$. Although a residual excess of cluster-associated galaxies, likely including infalling and backsplash galaxies, persists at these distances, the selection beyond $1.5\,R_{500c}$ is dominated by field interlopers. We therefore kept as cluster members only the selected galaxies within $1.5\,R_{500c}$. We note that a more refined membership probability is provided by the \texttt{RICH-CL} processing function, based on the approach presented by \citet{castignani_benoist}, from the \Euclid LE3 official cluster detection and characterization pipeline, as presented by \citet{bhargava_ec}. 

\begin{figure}
    \centering
    	\includegraphics[width=.85\linewidth]{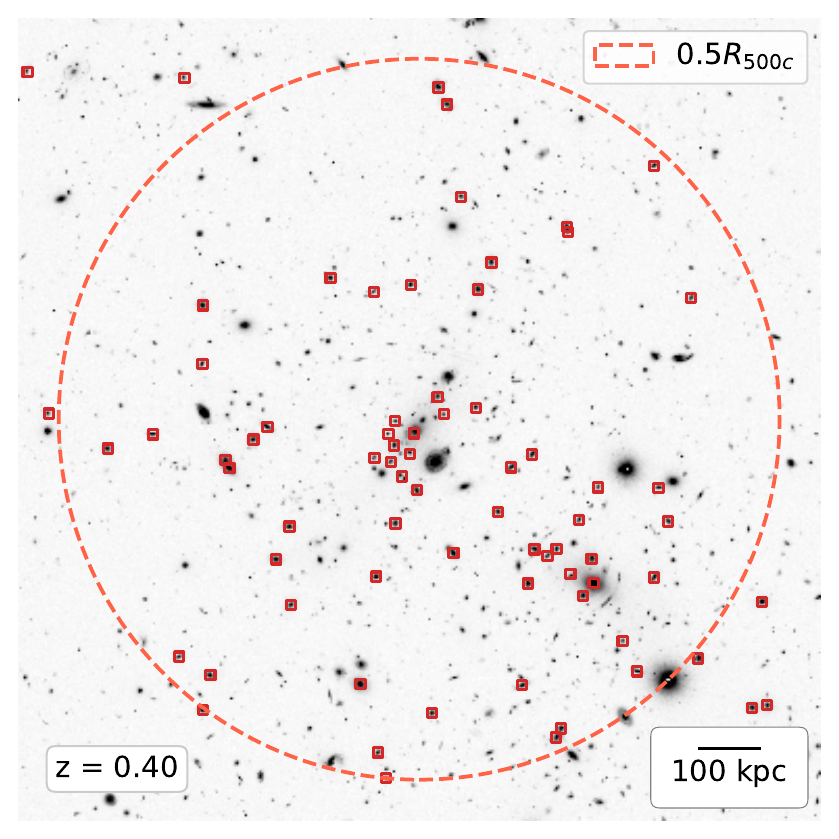}
        
    \caption{VIS cutout of $\ang{;4;} \times \ang{;4;}$ centered at the location of the cluster with most members (red squares), namely 1eRASS\,J041116.2$-$481850. Unselected galaxies are discarded either because they are not included in the visual morphology catalog or because they do not meet the selection or redshift membership criteria for the cluster. The cluster redshift, $0.5\,R_{500c}$ radius, and a physical scale bar are included in the figure. }
    \label{fig:cutouts_cl}
\end{figure}

Figure~\ref{fig:z_cl} shows the distribution of \Euclid cluster redshifts and the number of members assigned to each cluster in our sample. For our study of detailed morphology in clusters, we restricted our analysis to galaxies belonging to clusters with $0.2\leq z\leq 0.5$ to ensure adequate sample completeness across the stellar mass range (right panel of Fig.~\ref{fig:comp}). With a stellar mass completeness above the 50\% limit, we obtain a final sample of 1754 galaxies distributed across 71 clusters, corresponding to the unshaded redshift range shown in Fig.~\ref{fig:z_cl}. The cutout for the cluster with most members is displayed in Fig.~\ref{fig:cutouts_cl}, and Table~\ref{tab:selection} summarizes the sample selection steps performed.


\subsection{Morphology classification}\label{sec:morph_class}

\begin{figure}
    \centering
    	\includegraphics[width=.8\linewidth]{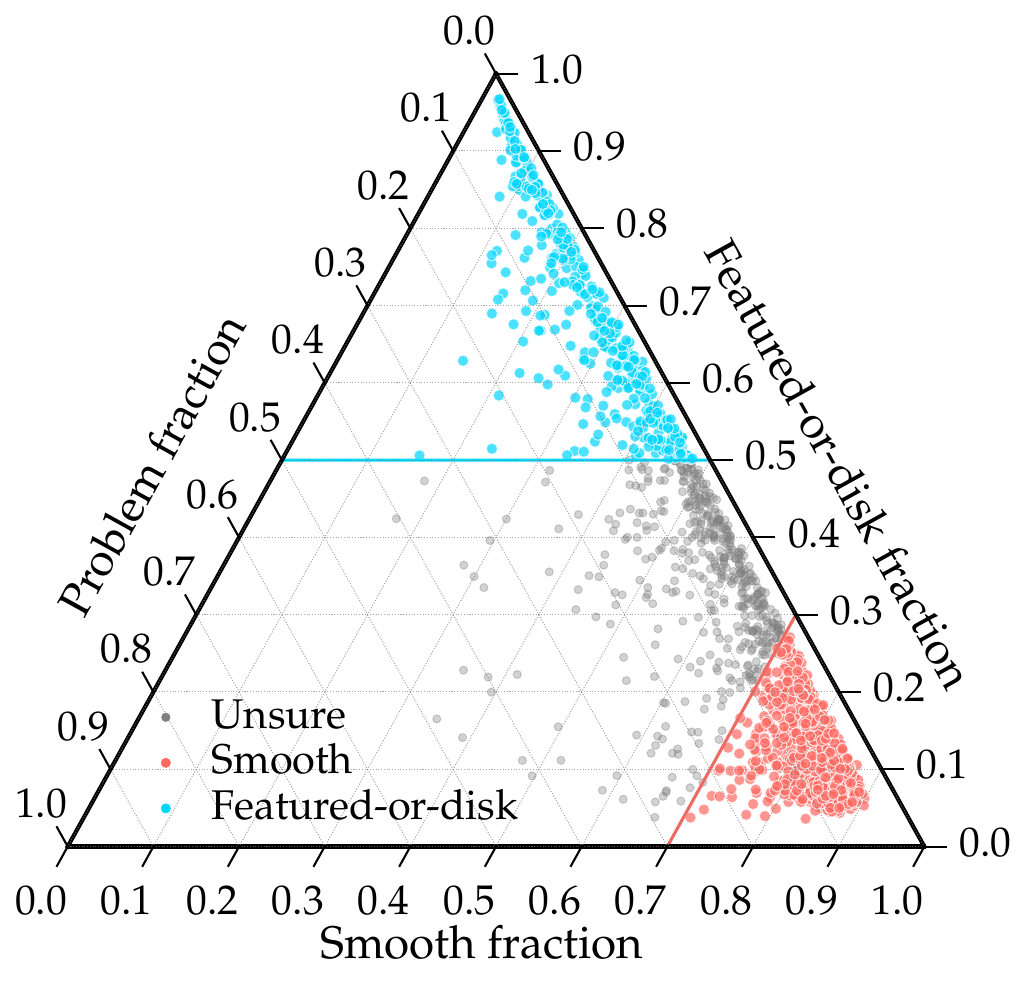}

    \caption{Analysis of the \texttt{Zoobot} expected vote fractions for the cluster members, presented as a ternary plot. Galaxies classified as smooth and featured-or-disk are superimposed in red and blue, respectively.}
    \label{fig:fractions_sc}
\end{figure}

To classify the identified galaxies into different morphology classes, we used the \Zoobot expected volunteer vote fractions, provided by the visual morphology catalog, together with the schema provided in the \texttt{zoobot.shared.schemas.euclid\_schema} Python module. We classified all the galaxies that meet the membership criteria out to $3\,R_{500c}$, so that the same morphological classification is available both for the cluster members within $1.5\,R_{500c}$ and for the population beyond this radius used to characterize the cluster outskirts (Sect.~\ref{sec:members}). Here, $p_{\rm A}$ corresponds to the expected vote fraction for a given answer, A, to a morphology question within the \Zoobot tree schema. We classified the galaxies as follows:

\begin{figure*}
    \centering
    	\includegraphics[width=.95\linewidth]{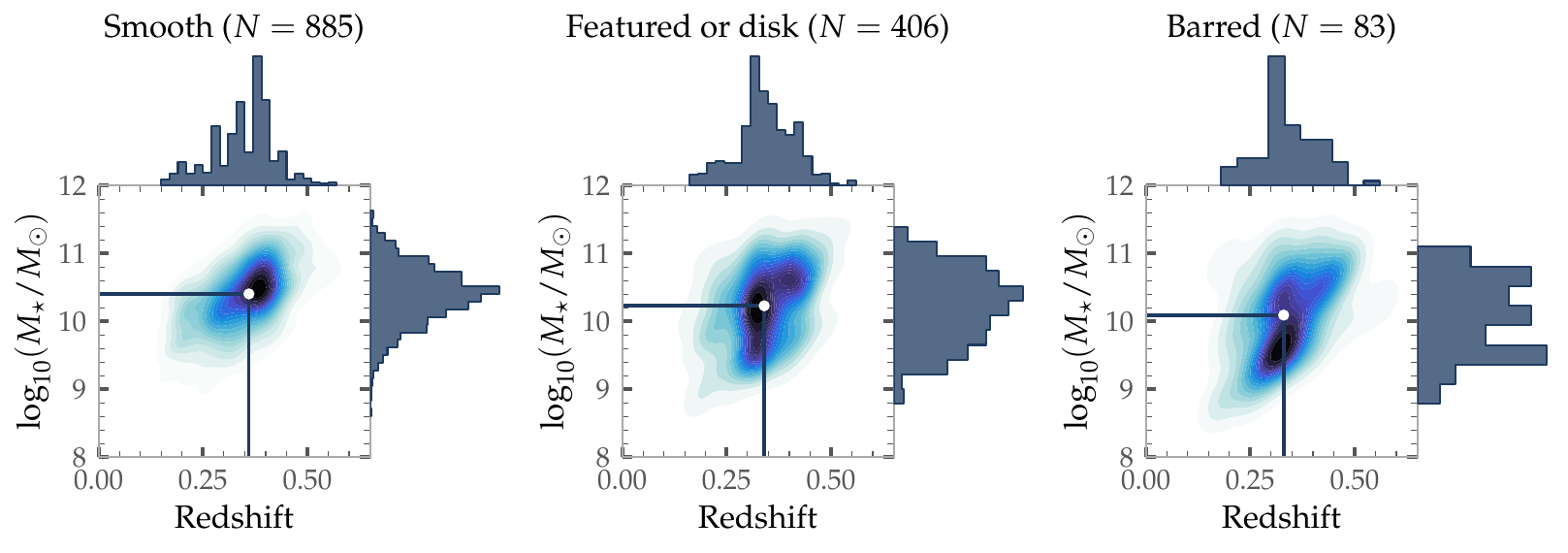}

    \caption{Distribution of stellar mass as a function of photometric redshift for different morphological classes in our sample of cluster members. The shaded contours represent 2D probability density estimated with a Gaussian kernel density estimator (darker colors indicate higher density). The marginal distributions are shown as histograms, and the solid lines indicate median values.}
    \label{fig:morph_dists}
\end{figure*}

\bi

    \item Smooth: $p_{\rm smooth} > 0.7$,

    \item Featured-or-disk: $p_{\rm featured-or-disc} > 0.5$,

    \item Barred: $p_{\rm featured-or-disc} > 0.5$ \& $p_{\rm edge-on}$ < 0.5 \& $p_{\rm bar}$ > 0.5.

\ei

We followed the same selection criteria for the selection of bars as \citet{euclid_marc}. The ternary plot displayed in Fig.~\ref{fig:fractions_sc} shows the distribution of the vote fractions corresponding to the ``smooth or featured?'' morphology question, for the cluster members classified as smooth and featured-or-disk. Galaxies that fall outside our classification schema are labeled as ``unsure'', since they are mostly poorly resolved galaxies with an uncertain classification (Fig.~\ref{fig:morph_unsure}). Random examples of galaxies corresponding to different morphological classes are displayed in Fig.~\ref{fig:morph_plots}, and Fig.~\ref{fig:best_bars} shows examples of cluster members with a high expected vote fraction to host a bar.

Our \Zoobot classification into smooth and featured-or-disk galaxies broadly corresponds to the separation between elliptical and disk morphologies. This is supported by \citet{quilley_ec}, who, using the same approach as in this work, recently showed that Q1 galaxies classified as smooth and featured-or-disk are consistent with an elliptical and disk classification derived from an elliptically symmetric 2D Sérsic profile fit \citep{sersic_1963}. We selected 0.7 as the threshold for $p_{\rm smooth}$ to classify smooth galaxies because, as discussed by \citet{quilley_ec}, this threshold has a significant impact on the resulting distribution, with increasing threshold values yielding more elliptical galaxies. However, as noted in previous work  \citep{willett_2013,simmons_2017,dominguez_2022,tobias_2025,smethurst_2025}, our selection of smooth galaxies could also include some featureless disk or S0 galaxies.


\section{Results and discussion}\label{sec:morph_results}

Figure~\ref{fig:morph_dists} shows the distribution of stellar mass as a function of redshift for the different morphological classes identified in our sample of cluster members. Using the \texttt{ndtest}\footnote{Written by Zhaozhou Li, \url{https://github.com/syrte/ndtest}} Python package, we performed 2D Kolmogorov--Smirnov (KS) tests \citep{kolmogorov33,smirnov48,peacock,fasano_test} to assess whether the smooth and featured-or-disk morphological classes share a common parent distribution in the stellar mass-redshift plane. The tests rejected the null hypothesis at high significance ($p$-value $\approx10^{-8}$), indicating that the two classes are significantly different in this plane. Smooth galaxies are concentrated in higher stellar masses, while featured-or-disk galaxies show a broader distribution and extend to lower masses, as illustrated by the marginal distributions. Barred galaxies, being a more specific subclass of featured-or-disk galaxies, have a smaller sample size ($N = 83$ vs. $N = 323$ unbarred featured-or-disk galaxies) and are found at lower stellar masses, with median $\log_{10}(M_\star/M_\odot)$ of 10.09 and 10.28, respectively. A Mann--Whitney U test \citep{mwtest} confirms that this difference is statistically significant ($p$-value $\approx\,10^{-3}$). To verify that this difference is not induced by a redshift mismatch between the barred and unbarred samples, we compared their redshift distributions. The two samples show no significant difference ($p$-value $= 0.75$ for a two-sample KS test), indicating that the lower stellar masses of barred galaxies are not a consequence of differing redshift distributions.

\begin{figure}
    \centering
    	\includegraphics[width=.8\linewidth]{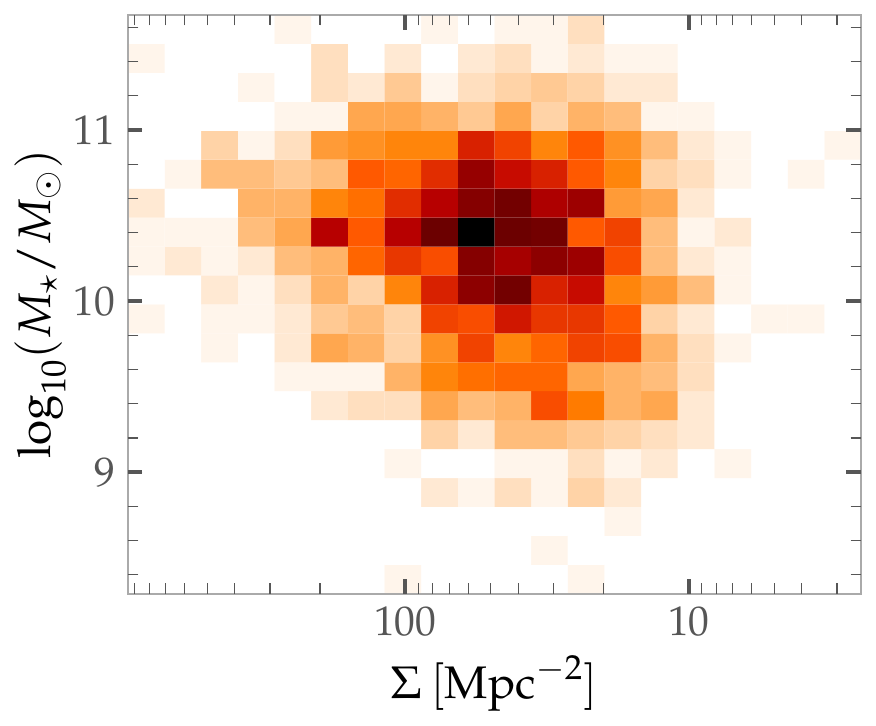}
    	\includegraphics[width=.8\linewidth]{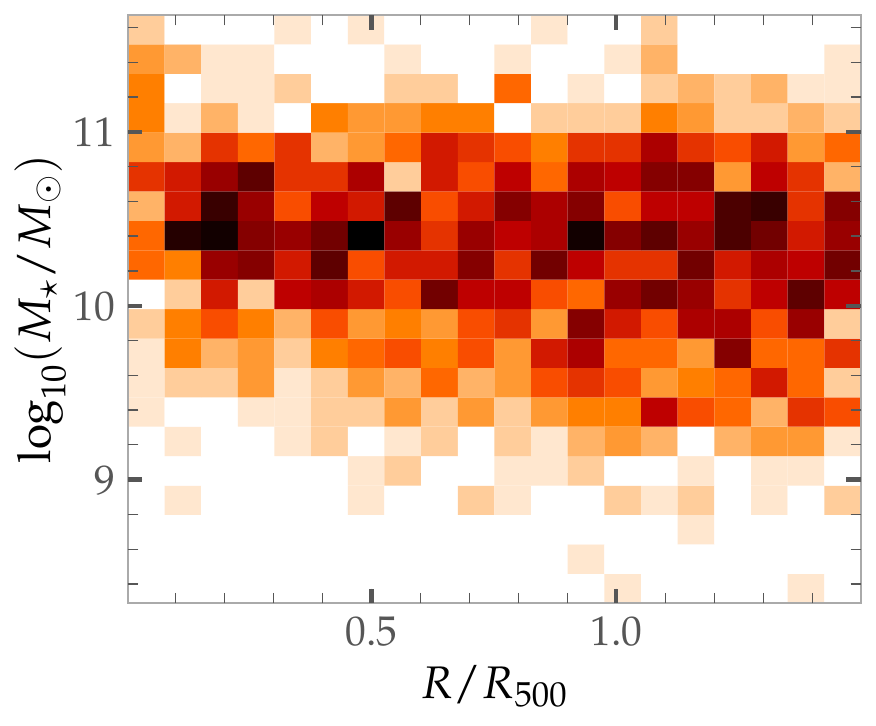}

    \caption{Stellar mass distribution of cluster members as a function of environment. \textit{Top panel}: Stellar mass as a function of local density. \textit{Bottom panel}: Stellar mass as a function of cluster-centric distance, normalized to $R_{\mathrm{500c}}$.}
    \label{fig:kde_members}
\end{figure}

Since local projected density and projected cluster-centric distance may play different roles for the evolution of galaxies of different morphologies  \citep[e.g., ][]{vulcani2023}, we included both parameters in our analysis. We consistently use projected measurements for the local density and cluster-centric distance but omit the word ``projected'' to reduce redundancy. The local density for the cluster members is computed as

\be
    \Sigma_N=\frac{N}{\pi\,D_N^{2}}\;,
\ee

\noindent where $N$ is the number of neighbors and $D_N$ is the projected distance to the $N^{\mathrm{th}}$ nearest neighbor. The neighbors were drawn from the MER catalog, applying a magnitude cut of $\IE<23$, the same photometric redshift reliability criteria as in Sect.~\ref{sec:euclid_Q1} and retaining only galaxies above the 50\% stellar mass completeness limit (Sect.~\ref{sec:members}). We used the average for $N=4$ and 5 \citep{baldry2006} to maintain consistency with previous work in this series of papers \citep{oryan_2026}. Figure~\ref{fig:kde_members} shows the distribution of galaxy stellar mass with both environment indicators for our final sample of cluster members.

\subsection{Smooth versus featured-or-disk galaxies}\label{sec:morph_fracs}

\begin{figure*}
    \sidecaption    	
    \includegraphics[width=12cm]{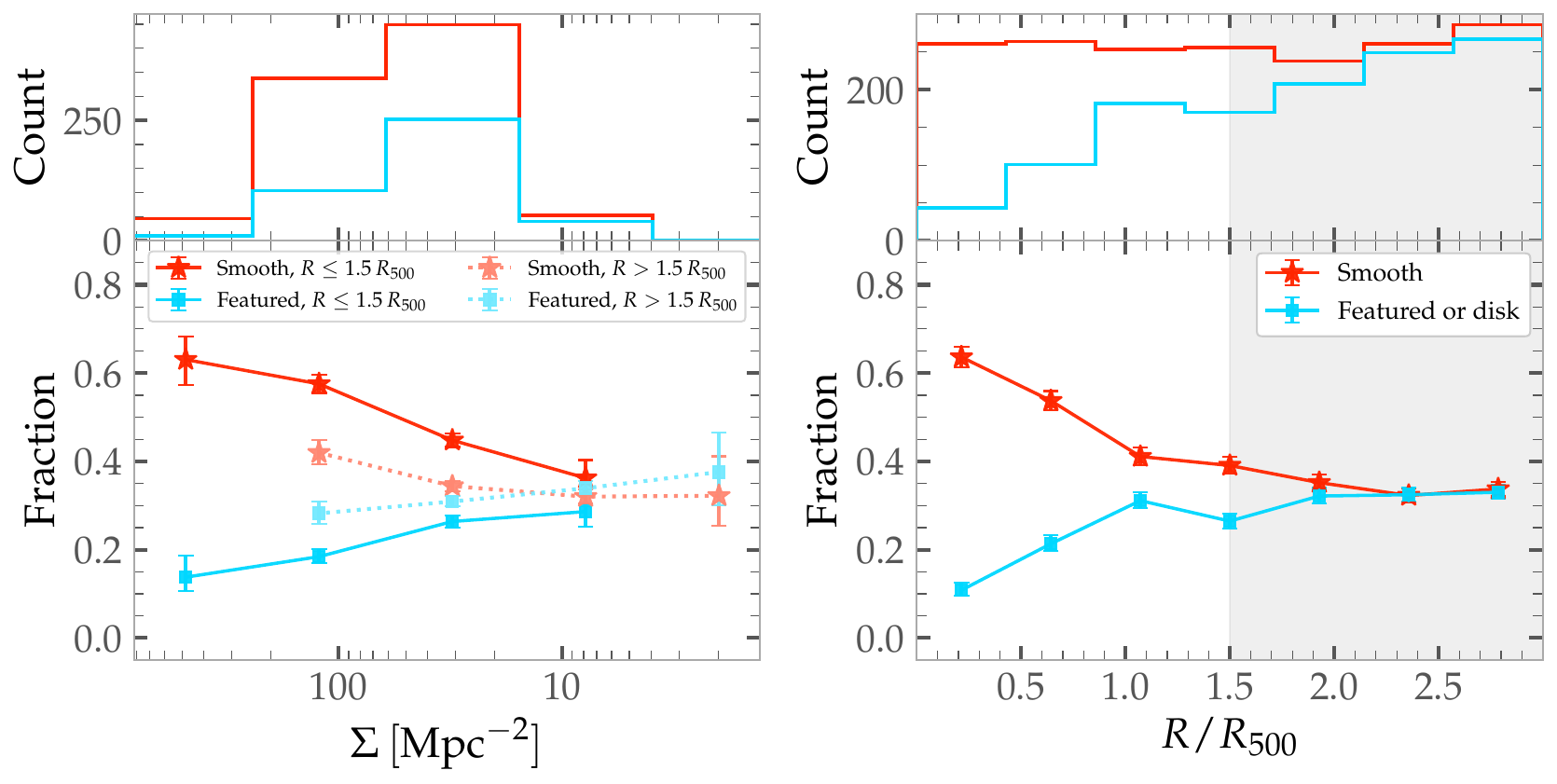}
    \caption{$T$--$\Sigma$ (\textit{left panel}) and $T$--$R$ (\textit{right panel}) relations for cluster members classified as smooth (red) and featured-or-disk (blue). \textit{Left panel}: Cluster members within $1.5\,R_{500c}$ (solid lines), together with galaxies (fainter, dotted lines) passing the photometric redshift membership selection beyond $1.5\,R_{500c}$, a mix of cluster outskirt galaxies and field interlopers (Sect.~\ref{sec:members}). \textit{Right panel}:  The gray-shaded region marks $R>1.5\,R_{500c}$, where the sample becomes increasingly dominated by field interlopers. The error bars represent the 68\% confidence interval under a beta-binomial posterior. Only bins with more than 15 galaxies in total are considered.}
    \label{fig:morph_fracs}
\end{figure*}

Figure~\ref{fig:morph_fracs} shows the change in the fraction of smooth and featured-or-disk galaxies as a function of both environment indicators. We accounted for selection effects by assigning each galaxy a weight of 1/completeness, with the completeness level as defined in Sect. \ref{sec:data}. Then, for each local density or cluster-centric distance bin, we computed the fraction of galaxies in each morphological class by dividing their weighted counts by the total weighted count of galaxies, including the galaxies classified as ``unsure''. We only considered bins with more than 15 galaxies in total.

The left panel of Fig.~\ref{fig:morph_fracs} presents the $T$--$\Sigma$ relation for the cluster members. Note that local density values are always inverted to facilitate analysis alongside cluster-centric distance, as the two quantities are anti-correlated. The figure separately illustrates the cluster members within $1.5\,R_{500c}$ and the galaxies that meet the membership criteria beyond this radius and up to $3\,R_{500c}$. These latter galaxies are dominated by a mixture of infalling and backsplash galaxies and field interlopers, as discussed in Sect.~\ref{sec:members}. For the cluster members, the smooth fraction decreases from $63_{-6}^{+5}\%$ at the denser regions to $36_{-4}^{+4}\%$ at the lower densities, while the featured-or-disk fraction shows the opposite trend, increasing from $14_{-3}^{+5}\%$ to $36_{-3}^{+4}\%$ over the same density range. We recover the well-established $T$--$\Sigma$ relation \citep[e.g., ][]{postman_2005,fasano_2015,vulcani2023} in a large sample of clusters at intermediate redshifts for the first time. In contrast, beyond $1.5\,R_{500c}$ the morphological segregation is markedly weaker, with a population which, at a given local density, is systematically more disk-rich compared with the interior of the cluster. This trend, with the fraction of featured-or-disk galaxies surpassing the fraction of smooth galaxies for the lowest densities, is fully consistent with the increased contamination of field interlopers measured in Fig.~\ref{fig:purity}, reinforcing our decision to restrict the cluster sample to $R\leq1.5\,R_{500c}$. The right panel of Fig.~\ref{fig:morph_fracs} displays the $T$--$R$ relation, with morphology fractions as a function of cluster-centric distance normalized to $R_{500c}$. The smooth fraction decreases from $64_{-2}^{+2}\%$ in the cluster cores to $39_{-2}^{+2}\%$ at $1.5\,R_{500c}$, while the featured-or-disk fraction increases correspondingly from $11_{-1}^{+2}\%$ to $26_{-2}^{+2}\%$. The $T$--$R$ relation is significantly stronger up to $\sim1\,R_{500c}$, in the innermost regions of the clusters, and flattens beyond it as we enter the cluster outskirts, a region where infalling and backsplash galaxies overlap with field interlopers that had largely been unexplored in previous studies \citep{vulcani2023}, with an increased contribution of field interlopers.

The decrease in the featured-or-disk fraction toward higher local densities and smaller cluster-centric distances is consistent with a loss of disk structure in galaxies driven by exposure to the cluster environment. Several physical processes contribute to this transformation, including ram-pressure stripping, tidal interactions, harassment, and strangulation \citep[e.g., ][]{boselli_2006}, each acting on different timescales and in different regions of the cluster. These environmental processes also lead to the suppression of star formation activity, a phenomenon known as ``environmental quenching,'' which has been found to precede morphological transformation for most galaxies in high-density environments \citep{gentile_ec,ghaffari_2026}. The overall effect is a progressive fading of disk features and a transformation to early-type morphologies, evolving backward along the Hubble sequence \citep{quilley_2022}, and the morphological segregation that we observe in Fig.~\ref{fig:morph_fracs}. The median local density within $0.5\,R_{500c}$ is $\Sigma=90$\,Mpc$^{-2}$, decreasing to $\Sigma=44$\,Mpc$^{-2}$ in the $0.5$--$1\,R_{500c}$ range, $\Sigma=28$\,Mpc$^{-2}$ between $1$ and $2\,R_{500c}$, and $\Sigma=21$\,Mpc$^{-2}$ beyond $2\,R_{500c}$. This steep density gradient, with the most pronounced drop occurring within $\sim R_{500c}$, is consistent with the transitions observed in both the $T$--$\Sigma$ and $T$--$R$ relations (Fig.~\ref{fig:morph_fracs}), and supports the interpretation that the dominant morphological transformation mechanisms operate most efficiently in the inner cluster regions where the intracluster medium density is highest.

Another way of quantifying the $T$--$\Sigma$ and $T$--$R$ relations in our sample of cluster members is to directly use the median of the smooth and featured-or-disk expected vote fractions for each of the environment bins. In this way, we do not depend on the prior definition of different classes but simply analyze how the smoothness of the sample evolves. This is illustrated in Fig.~\ref{fig:smooth_feat_votes}, recovering the same trends discussed above.

\begin{figure*}
    \sidecaption
    \parbox[b]{12cm}{%
        \includegraphics[width=12cm]{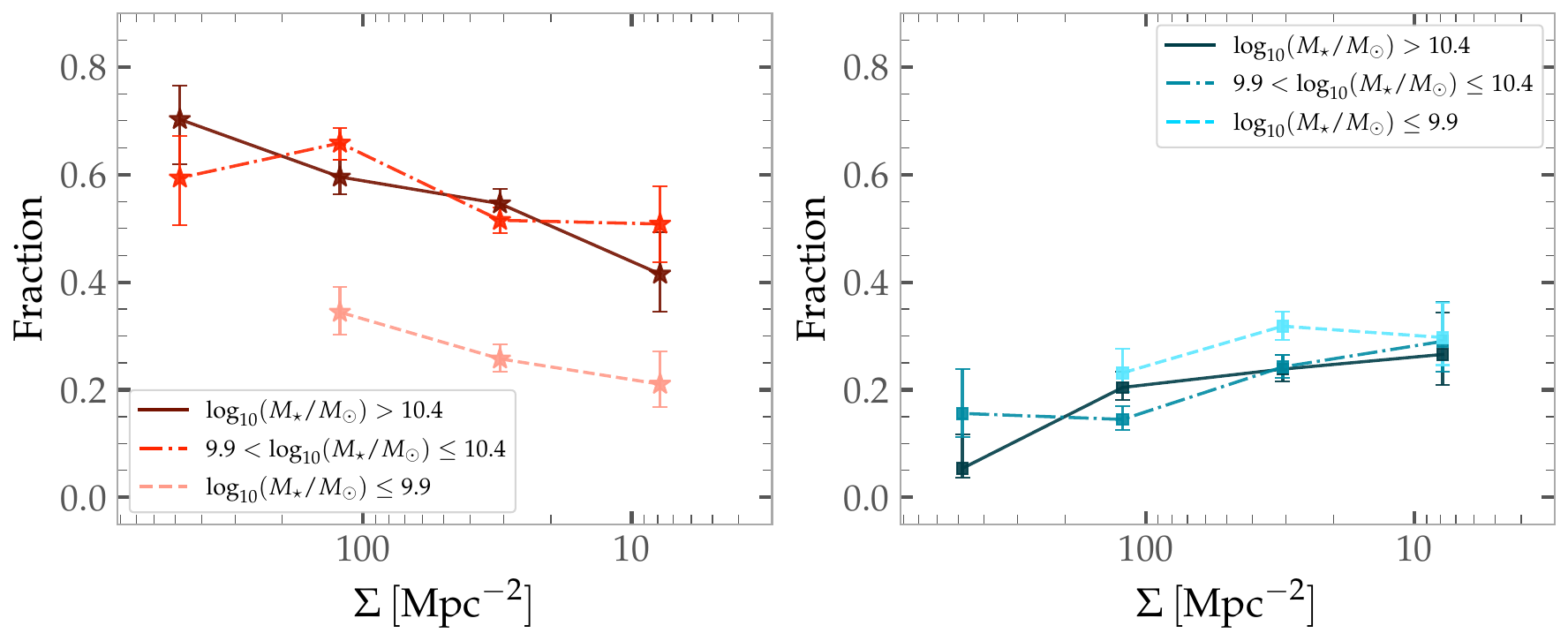}\\
        \includegraphics[width=12cm]{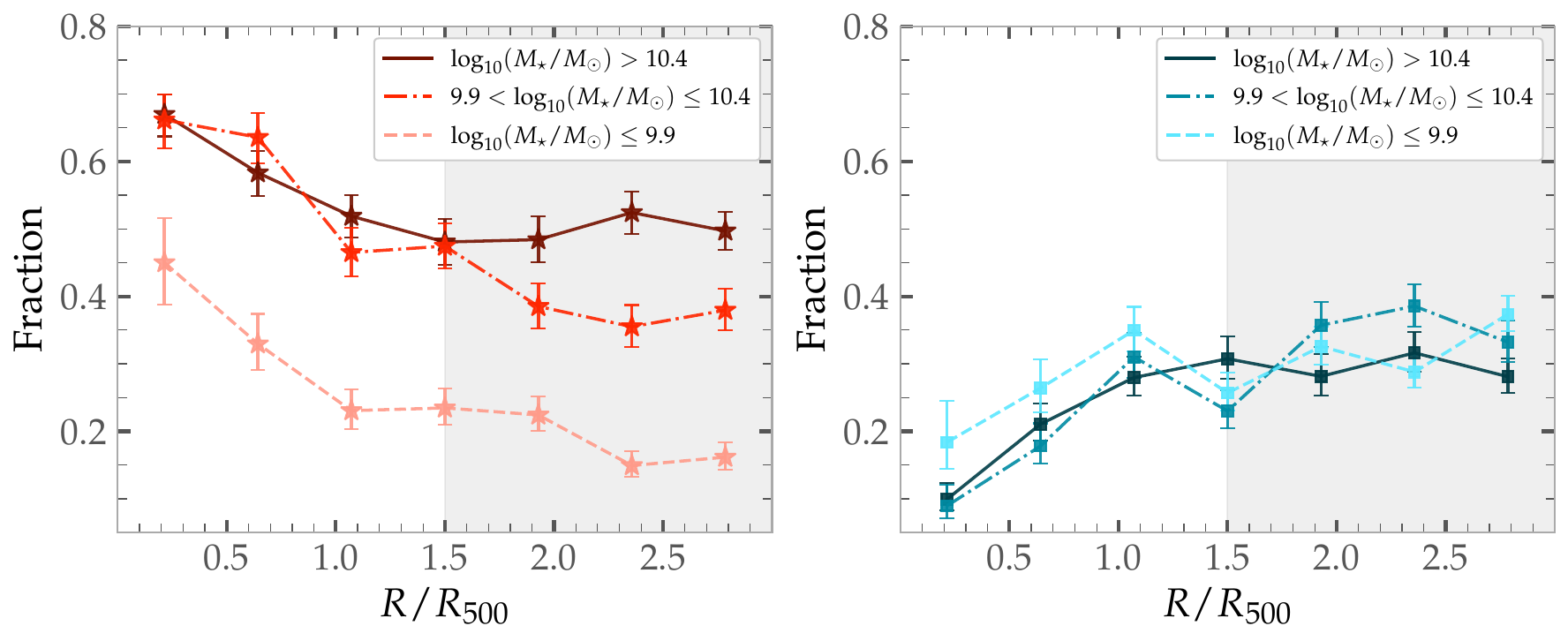}}
    \caption{$T$--$\Sigma$ (\textit{upper panels}) and $T$--$R$ (\textit{bottom panels}) relations for the cluster members classified as smooth (red) and featured-or-disk (blue) in bins of stellar mass. The $T$--$\Sigma$ relation is measured using only the cluster members (within $1.5\,R_{500c}$). The lighter and darker colors represent bins of lower and higher stellar mass, respectively. The gray-shaded region in the bottom panels is described in Fig.\,\ref{fig:morph_fracs}.}
    \label{fig:morph_fracs_bin}
\end{figure*}

\begin{figure*}
    \sidecaption
    	\includegraphics[width=12cm]{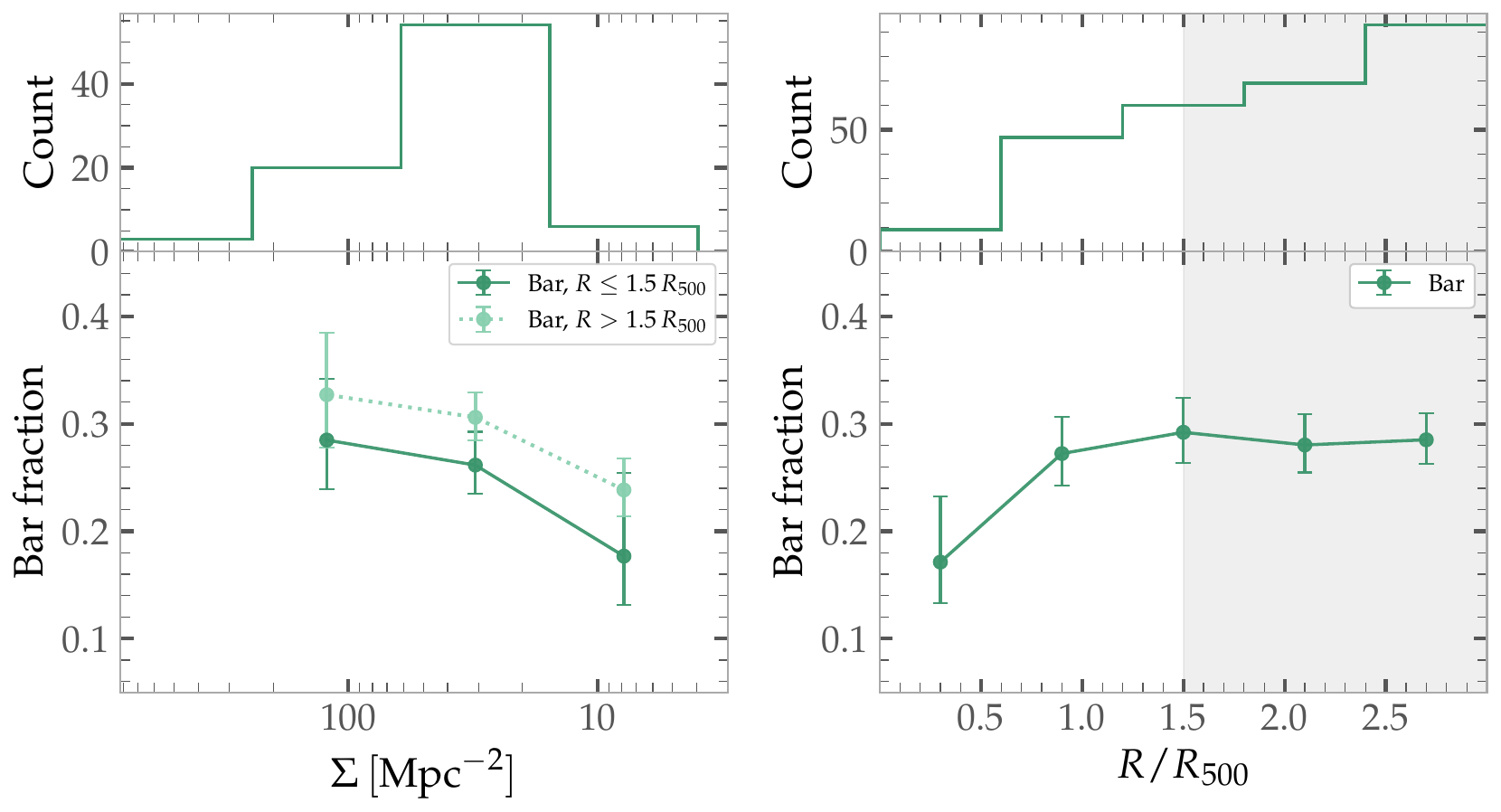}
    \caption{Same as in Fig.~\ref{fig:morph_fracs} but for the $f_{\rm bar}$--$\Sigma$ and $f_{\rm bar}$--$R$ relations. Here, the bar fraction is computed relative to the total number of featured-or-disk non edge-on galaxies. }
    \label{fig:bar}
\end{figure*}

Figure~\ref{fig:morph_fracs_bin} shows the $T$--$\Sigma$ and $T$--$R$ relations for different bins of stellar mass, defined using the 0.33 and 0.66 quantiles of the distribution. Here, the $T$--$\Sigma$ is measured taking only into account the cluster members, while the $T$--$R$ is shown for the full distance range. The persistence of the relations across all stellar mass bins supports the existence of an environmental effect that acts independently of mass-driven evolution, in agreement with previous works that have separated the role of stellar mass and environment in shaping galaxy evolution \citep[e.g., ][]{bamford_2009, peng_2010}. The fact that the slope of the $T$--$R$ relation for smooth galaxies is steeper in the low-mass bin than for higher masses further suggests that this environmental effect is more significant for low-mass cluster members, which are more easily affected by the cluster environment \citep{contini_2020}. In contrast, the high-mass bins show consistently higher smooth fractions across the entire range of environments, indicating that the fraction of smooth galaxies increases with increasing stellar mass regardless of environment. The fraction of smooth galaxies is considerably reduced in the less massive bin, while the fraction of featured-or-disk galaxies with environment is similar for all stellar mass bins. This is due to the number of smooth cluster members being reduced for the less massive bin, while the number of cluster members classified as ``unsure'' increases, keeping the fractions of featured-or-disk cluster members similar.

Directly comparing the fractions found in this work with values reported in the literature is extremely difficult given the different sample selections, stellar mass and redshift bins, morphological classification methods, environmental indicators (e.g., the value of $N$ adopted to measure $\Sigma_N$), or data used. Many studies have examined the $T$--$\Sigma$ relation in clusters in the local Universe \citep[e.g., ][]{fasano_2015,houghton_2015,vulcani2023} and at higher redshifts \citep[e.g., ][]{postman_2005,smith_2005,poggianti_2008}. Recently, \citet{cleland_ec} used the morphological classification of \citet{quilley_ec}, which separates ETGs and LTGs in the Sérsic index $n$ versus $u-r$ color plane, to measure the evolution of the ETG fraction as a function of local density in different redshift and stellar mass bins. Having to account for their different sample selection and morphological classification, and taking all their identified ETGs to be elliptical galaxies, the trend observed for their lowest redshift ($0.25<z<0.5$) bin is consistent with this work. Moreover, as shown in Fig.~\ref{fig:smooth_feat_dens7}, these results are not significantly changed if we use $N=7$ to compute $\Sigma_N$, as in \citet{cleland_ec}. Overall, the environmental trends on smooth and featured-or-disk galaxies presented in this paper agree with the existing literature mentioned above.


\subsection{Barred galaxies}\label{sec:bars}


\begin{figure}
    \centering
    	\includegraphics[width=.85\linewidth]{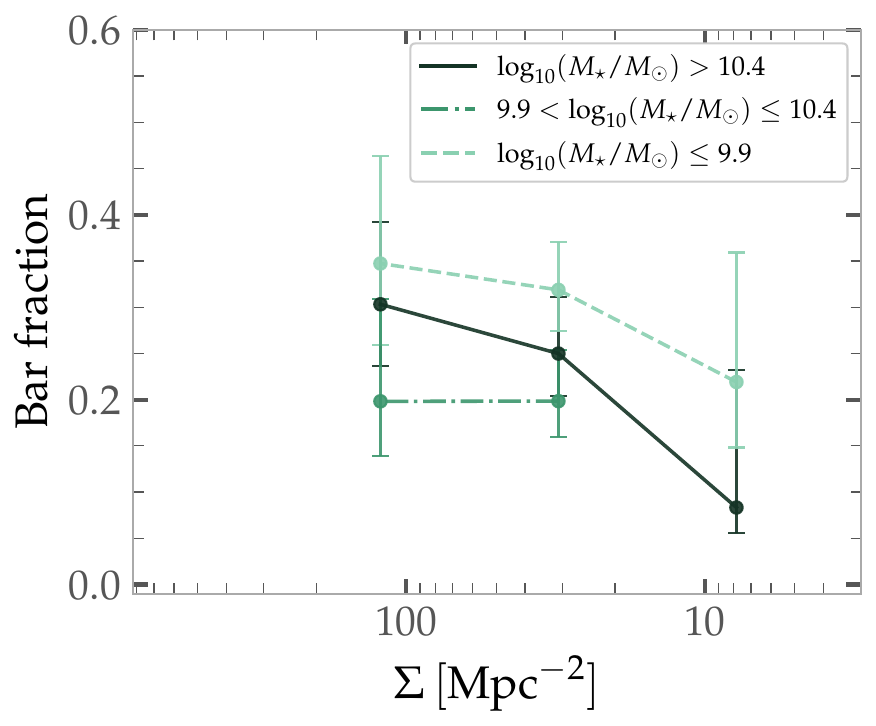}

    \caption{$f_{\rm bar}$--$\Sigma$ relation in bins of stellar mass, measured using only the cluster members (within $1.5\,R_{500c}$).}
    \label{fig:bar_mbins}
\end{figure}

The bars act as major internal drivers of the evolution of disk galaxies, as they fuel nuclear star formation by funneling gas toward the center of the galaxy \citep{kormendy_2004}, and redistribute angular momentum within the galaxy \citep{Athanassoula_2013}. Although it has been thoroughly studied in the literature, to date there is no clear consensus on the effects of the environment on bar formation and evolution \citep[e.g., ][]{aguerri_2009,abreu_2010,barway_2011,castignani_2022,aguerri_2023}. The role of the environment in this picture is complex. While the formation of bars can be triggered by tidal interactions   \citep{martinez_val_2017, peschken_2019}, high-density environments can also suppress or even destroy bars, depending on the internal properties of the galaxy, such as its stellar mass, and the intensity of the interaction \citep{abreu_2012}.

We computed the bar fraction as
\be    
f_{\mathrm{bar}} = \frac{N_{\mathrm{bar}}}{N_{\mathrm{gal}}}\;,
\ee
where $N_{\mathrm{bar}}$ is computed using the selection criteria defined in Sect.~\ref{sec:morph_class} and $N_{\mathrm{gal}}$ is defined as the number of featured-or-disk galaxies, excluding edge-on galaxies ($p_{\rm edge-on} < 0.5$), which leaves $N_{\rm gal}=324$ of the 406 featured-or-disk cluster members. Hereafter, we refer to the relations between the bar fraction and local density and cluster-centric distance as $f_{\rm bar}$--$\Sigma$ and $f_{\rm bar}$--$R$, respectively. The average bar fraction in our sample of cluster members, $f_{\rm bar}\approx26\%$, is consistent with the values reported by \citet{euclid_marc} for the same redshift range using the entire Euclid Q1 visual morphology catalog and bar classification schema, although direct comparison is challenging given their more stringent selection criteria ($\IE<20.5$) and, therefore, different sample completeness. This suggests that, at $0.2 \le z \le 0.5$, the cluster environment does not produce a strong global enhancement or suppression of the bar fraction.

Figure~\ref{fig:bar} shows the $f_{\rm bar}$--$\Sigma$ and $f_{\rm bar}$--$R$ relations in our sample. Note that the number of bins is reduced to improve statistical significance. For both the cluster members within $1.5\,R_{500c}$ and the cluster outskirts beyond it, the $f_{\rm bar}$--$\Sigma$ relation is essentially flat in the two intermediate-density bins and shows a mild decrease at lower densities, suggesting a decreased presence of barred galaxies in very low-density regions. Beyond $1.5\,R_{500c}$, the bar fraction lies systematically above the cluster value at any given local density. This is explained by the slight suppression of the bar fraction in the cluster core shown by the $f_{\rm bar}$--$R$ relation, which may reflect the heating of the disk component in the densest regions, inhibiting bar formation \citep{abreu_2010, abreu_2012}, although the small number of featured-or-disk non-edge-on galaxies in the innermost bin requires caution before drawing strong conclusions from this point alone.

The breakdown by stellar mass displayed in Fig.~\ref{fig:bar_mbins} reveals that this picture is mass-dependent, with a more pronounced decrease in the bar fraction toward lower densities for the highest-mass galaxies. Given the large uncertainties and the overlap, these density trends should be regarded as tentative, but they suggest that the mild decline of the bar fraction toward very-low densities seen in Fig.~\ref{fig:bar} is driven mainly by the most massive galaxies. This mass-dependent behavior suggests that, although bar formation is primarily driven by internal disk instabilities \citep{kormendy_2004}, the cluster environment may play a complementary role that depends on the physical properties of the host galaxy. Figure~\ref{fig:bar_mbins} also shows a nonmonotonic dependence of the bar fraction on stellar mass. At fixed local density, galaxies with $9.9<\log_{10}(M_\star/M_\odot)\leq10.4$ show the lowest bar fraction, $f_{\rm bar}\approx0.2$, whereas both lower- and higher-mass galaxies reach $f_{\rm bar}\approx0.3$--$0.35$ in the densest environments. Although the differences between stellar mass bins are within the uncertainties, this behavior resembles the bimodal distribution reported for low-redshift galaxies, with a minimum at $\log_{10}(M_\star/M_\odot)\approx10.2$ \citep{nair2010,masters2012,mukundan2025}, which suggests that it is already in place in cluster galaxies at $0.2\leq z\leq0.5$.

The measured bar fraction as a function of stellar mass is, however, sensitive to observational and classification biases. Bar length scales with galaxy size and stellar mass, so the short bars typical of low-mass disks can remain undetected when they are not well resolved \citep{erwin_2018}. With a full width at half maximum of \ang{;;0.16} for the VIS point spread function, corresponding to $\simeq$0.55--1\,kpc over our redshift range, bars with semi-major axes below $\simeq$1.1--2\,kpc may not be reliably detected, particularly at $z\sim0.5$. In addition, strong bars (longer and obvious) are more frequent in massive galaxies, while weak bars (smaller and fainter) dominate lower-mass galaxies \citep{tobias_2021}, and visual classifications, such as those used to train \Zoobot, preferentially detect stronger bars. Both effects would lower the measured bar fraction of low-mass galaxies, so the higher $f_{\rm bar}$ we find in this regime relative to more massive galaxies is likely a lower limit. Future \Euclid data releases, with substantially larger samples of cluster members, will be key to confirming these trends and reducing the statistical uncertainties that currently limit their interpretation.


\section{Conclusions}\label{sec:conclusions}

This paper harnesses the unique characteristics of the \Euclid space mission to study the effects of cluster environment on detailed visual galaxy morphology at intermediate redshifts. We measured the $T$--$\Sigma$ and $T$--$R$ relations in a sample of 1754 cluster members distributed across 71 clusters at $0.2\leq z\leq 0.5$, and extended them to the cluster outskirts, up to $3\,R_{500c}$ ($\sim2\,R_{200c}$). This complements the work by \citet{oryan_2026}, who created a large compilation of HST galaxy morphologies in clusters within a similar redshift range. Using the Euclid Q1 visual morphology catalog, built by fine-tuning the \texttt{Zoobot} deep foundation model with dedicated labels for \Euclid galaxies, we classified the identified galaxies as smooth or featured-or-disk, with additional criteria to also identify barred galaxies. The morphology fractions in each bin of projected local density and cluster-centric distance were corrected for stellar mass completeness by weighting each galaxy by the inverse of its completeness level. Our main results can be summarized as follows.

\bi
\setlength\itemsep{.1em}

    \item We recovered the $T$--$\Sigma$ relation in a statistically significant sample of intermediate-redshift clusters, with a stronger influence of the cluster environment on galaxy morphology in the densest regions. For the cluster members within $1.5\,R_{500c}$, the smooth fraction decreases from $63_{-6}^{+5}\%$ at the highest densities to $36_{-4}^{+4}\%$ at lower densities, while the featured-or-disk fraction increases from $14_{-3}^{+5}\%$ to $36_{-3}^{+4}\%$ over the same density range. Beyond $1.5\,R_{500c}$, the morphological segregation is weaker, with the fraction of featured-or-disk galaxies surpassing the fraction of smooth galaxies for the lowest densities.
    
    \item We measured the $T$--$R$ relation up to $3\,R_{500c}$, with the smooth fraction decreasing from $64_{-2}^{+2}\%$ in the cluster cores to $39_{-2}^{+2}\%$ at $1.5\,R_{500c}$, and the featured-or-disk fraction increasing correspondingly from $11_{-1}^{+2}\%$ to $26_{-2}^{+2}\%$. The relation is significantly stronger within $\sim R_{500c}$ and flattens beyond it, as we enter the cluster outskirts, where infalling and backsplash galaxies are mixed with field interlopers.
    
    \item Both relations persist across all the stellar mass ranges studied, indicating an environmental effect that acts independently of mass-driven evolution. Higher-mass galaxies show consistently higher smooth fractions at nearly all local densities and cluster-centric distances, while the steeper $T$--$R$ slope for low-mass smooth galaxies suggests that the environmental effect is more significant for the low-mass population, which is more easily affected by the cluster environment.
    
    \item The bar fraction relative to featured-or-disk non-edge-on galaxies is $\approx26\%$, consistent with the value reported by \citet{euclid_marc} for the full Euclid Q1 visual morphology catalog, indicating that the cluster environment does not produce a strong global enhancement or suppression of the bar fraction at $0.2\leq z\leq0.5$. Within the cluster, the density trends of the bar fraction are weak and remain tentative, with the strongest decline toward lower densities observed for the most massive galaxies. Beyond $1.5\,R_{500c}$, the bar fraction lies systematically above the cluster value, consistent with the slight suppression of bars observed in the cluster core.
\ei

Overall, these results draw a consistent physical picture of morphological transformation driven by the cluster environment. The morphological segregation traced by the $T$--$\Sigma$ and $T$--$R$ relations reflects a progressive loss of disk structure in galaxies exposed to the dense cluster environment, which acts through a combination of physical processes such as ram-pressure stripping and tidal interactions. The steep local density gradient in our sample, dropping from $\Sigma=90$\,Mpc$^{-2}$ within $0.5\,R_{500c}$ to $\Sigma=21$\,Mpc$^{-2}$ beyond $2\,R_{500c}$, with the most pronounced decrease occurring within $\sim R_{500c}$, mirrors the transitions observed in both environmental relations and supports a scenario in which these transformation mechanisms operate most efficiently in the innermost cluster regions, where the intracluster medium is densest. The mass dependence of the bar fraction further suggests that the impact of the environment may be modulated by the internal properties of the host galaxies, with the slight suppression of bars in the cluster cores possibly reflecting the heating of the disk component in the densest regions.

This work demonstrates that the unique combination of high spatial resolution and sensitivity, with a large survey area, provided by \Euclid is key to improving our understanding of the effects of cluster environments on the visual morphology of galaxies at $0.2\leq z\leq 0.5$. The final EWS will provide detailed visual morphology measurements for around $10^{8}$ galaxies up to $z\sim1.5$, enabling comprehensive studies of environmental effects on galaxy morphology over cosmic time.


\begin{acknowledgements}

PMB acknowledges the support of the ESA science faculty funding for this project hosted at the Centro de Astrobiología (CAB) in CSIC-INTA.
\AckQone
\AckDatalabs
\AckEC
We made extensive use of Python throughout the entire process, including the packages \texttt{pandas}\footnote{\url{https://github.com/pandas-dev/pandas}}, \texttt{seaborn} \citep{seaborn}, \texttt{numpy} \citep{numpy}, \texttt{matplotlib} \citep{matplotlib}, and \texttt{astropy} \citep{astropy}.

\end{acknowledgements}


\bibliographystyle{aa} 
\bibliography{Bibliography} 

\begin{appendix}
  \onecolumn 

\section{Morphology cutouts}
\label{app_morph}

\begin{figure}[h!]
    \centering
    	\includegraphics[width=\linewidth]{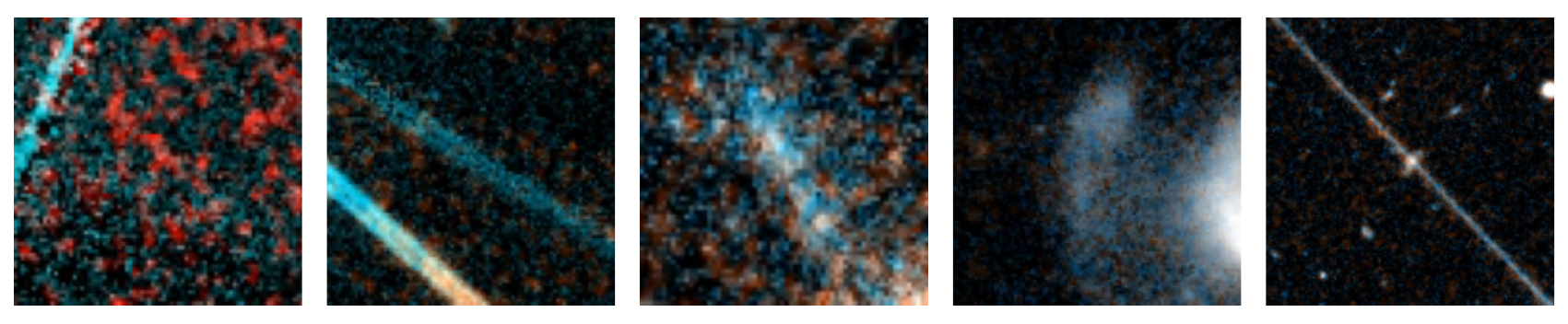}

    \caption{Examples of objects likely to be artifacts.}
    \label{fig:prob}
\end{figure}

\begin{figure}[h!]
    \centering
    	\includegraphics[width=\linewidth]{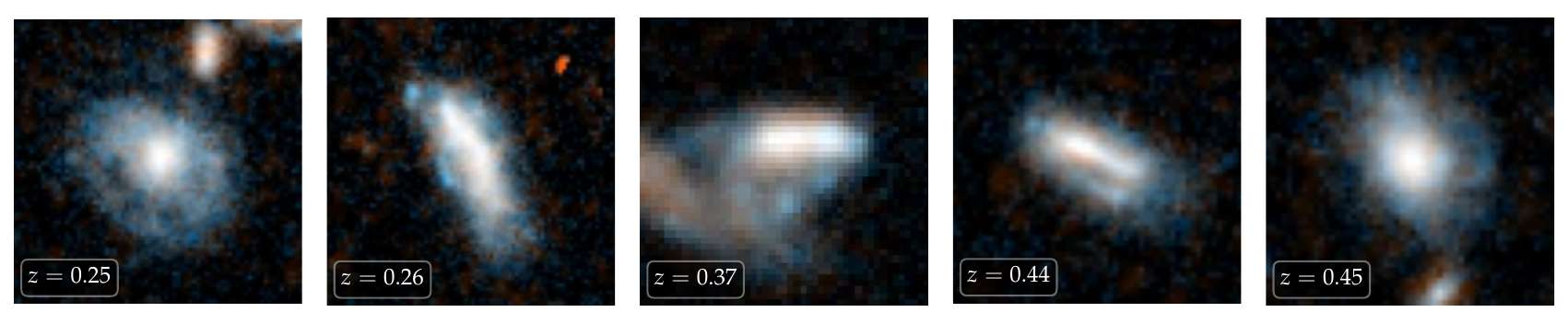}

    \caption{Examples of galaxies classified as ``unsure.''}
    \label{fig:morph_unsure}
\end{figure}

\begin{figure}[h!]
    \centering
    	\includegraphics[width=\linewidth]{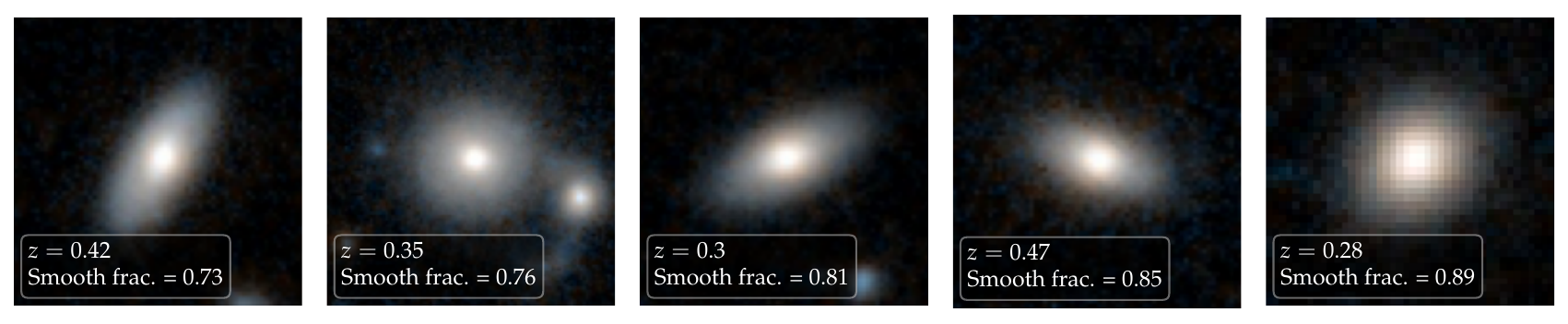}
        
    	\includegraphics[width=\linewidth]{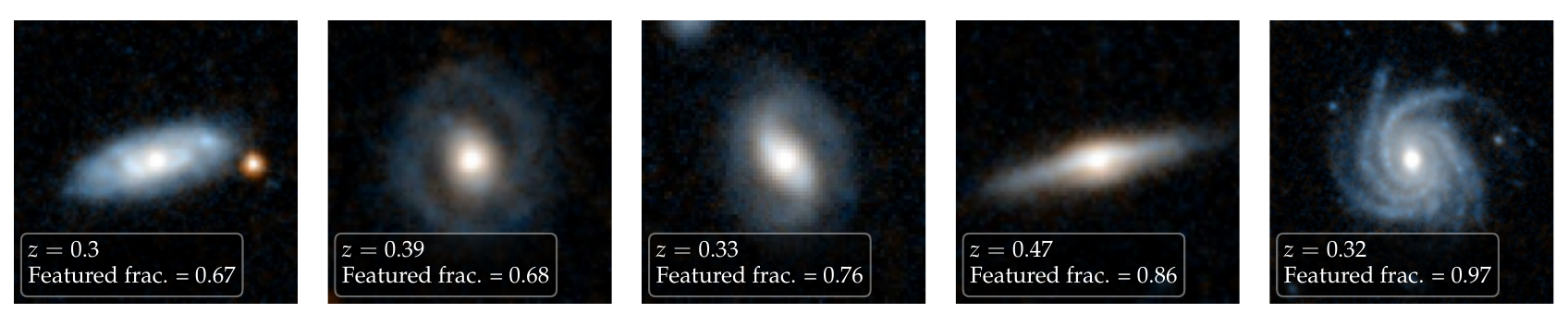}
        
    	\includegraphics[width=\linewidth]{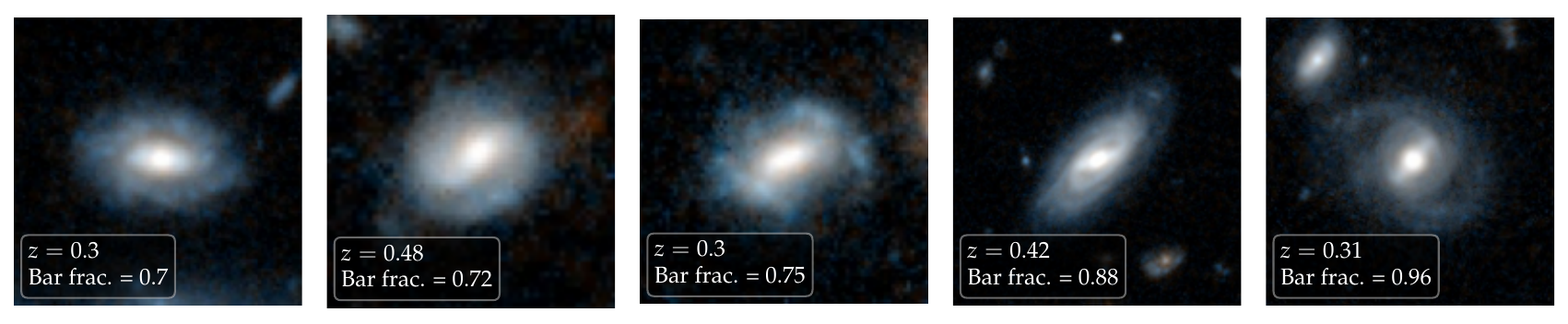}
        
        \includegraphics[width=\linewidth]{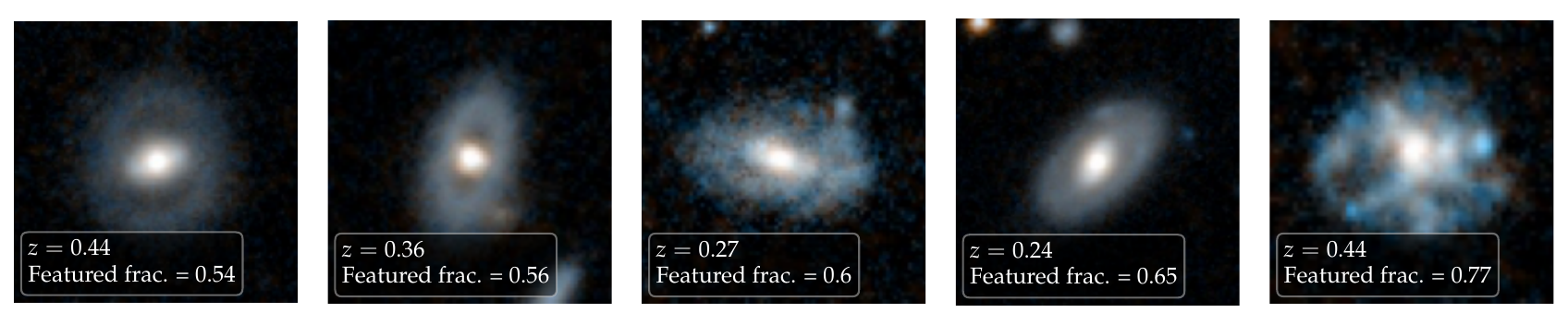}
        
        \includegraphics[width=\linewidth]{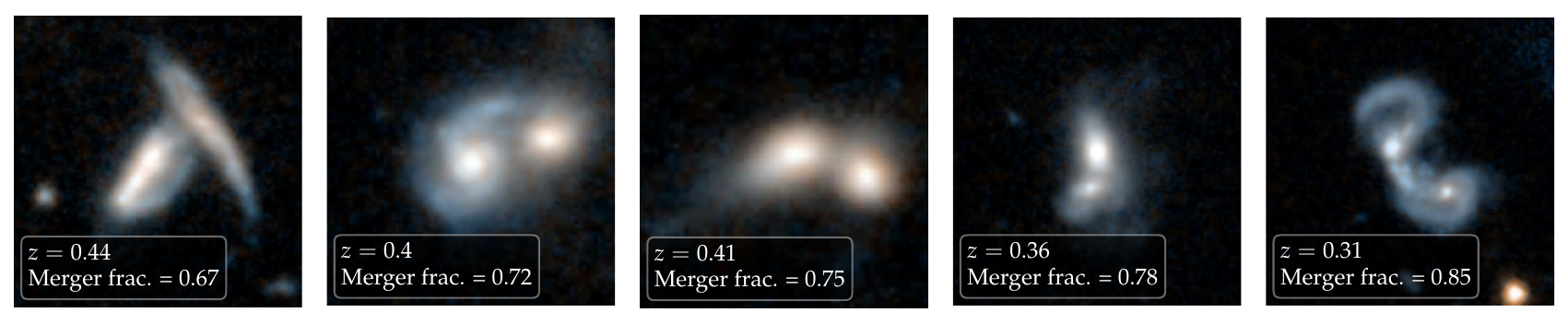}
        
    \caption{Mosaic of random galaxy examples corresponding to different morphologies. From top to bottom: Smooth, featured-or-disk, barred, featured-or-disk and $p_{\mathrm{has\_spiral\_arms}}<0.5$, and merging galaxies.}
    \label{fig:morph_plots}
\end{figure}

\begin{figure}[h!]
    \centering
    	\includegraphics[width=\linewidth]{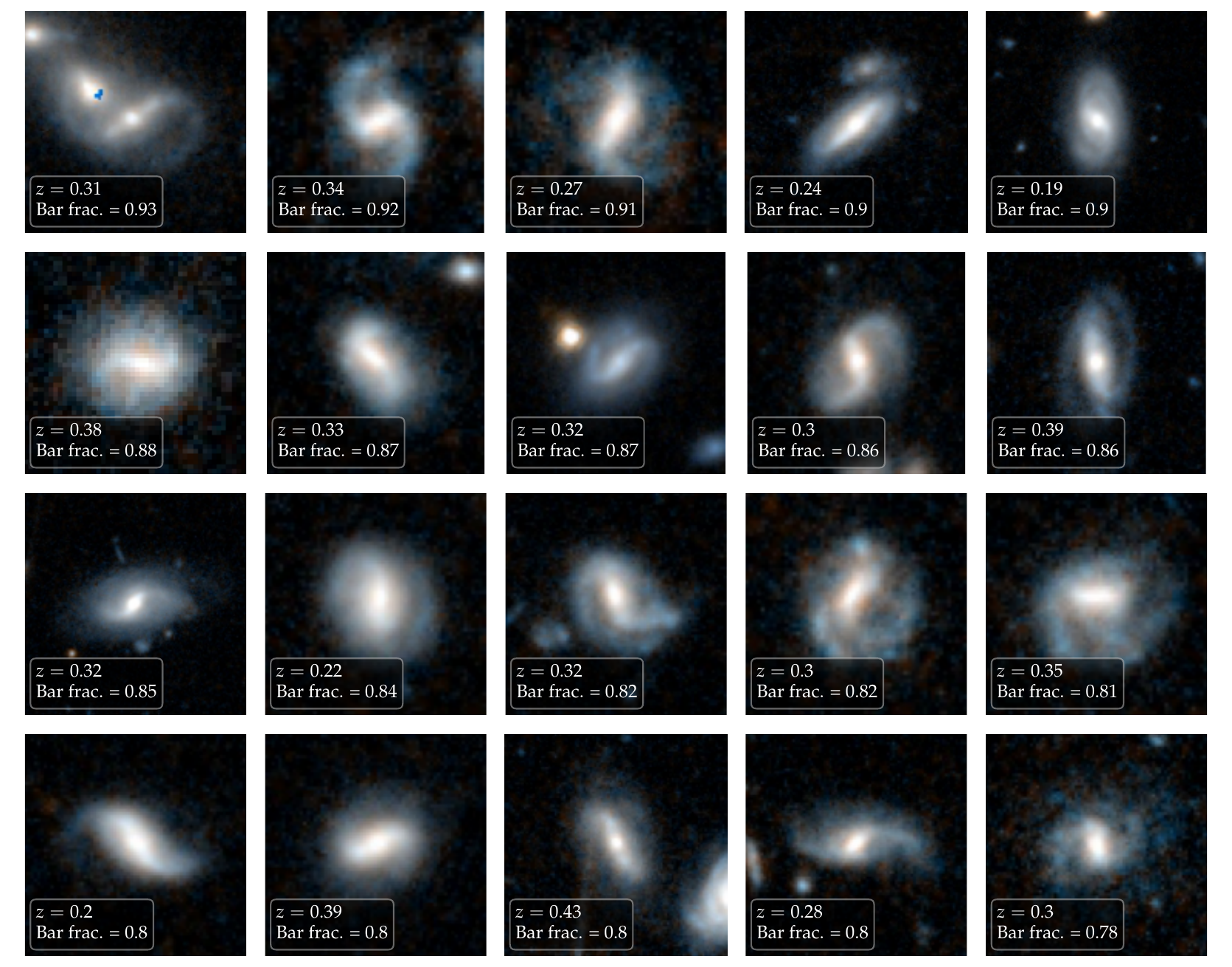}

    \caption{Mosaic of the cluster members with the highest expected vote fraction to host a bar.}
    \label{fig:best_bars}
\end{figure}

Example images of representative sample galaxies. Figure \ref{fig:prob} shows examples of objects that were removed from our sample as they are likely to be artifacts, given that they have a \Zoobot expected volunteer vote fraction of \texttt{smooth-or-featured\_problem\_fraction} $> 0.5$. In Fig.~\ref{fig:morph_unsure}, we display examples of galaxies labeled as ``unsure'' at different redshifts. Figure~\ref{fig:morph_plots} displays examples of galaxies classified as smooth, featured-or-disk, barred, featured-or-disk  with $p_{\mathrm{has\_spiral\_arms}}<0.5$, and mergers at different redshifts. A more detailed view of the galaxies with the highest expected probability of hosting a bar is presented in Fig.~\ref{fig:best_bars}.

\FloatBarrier
\section{Additional plots}
\label{app_add}

\begin{figure}[h!]
    \sidecaption
    \includegraphics[width=12cm]{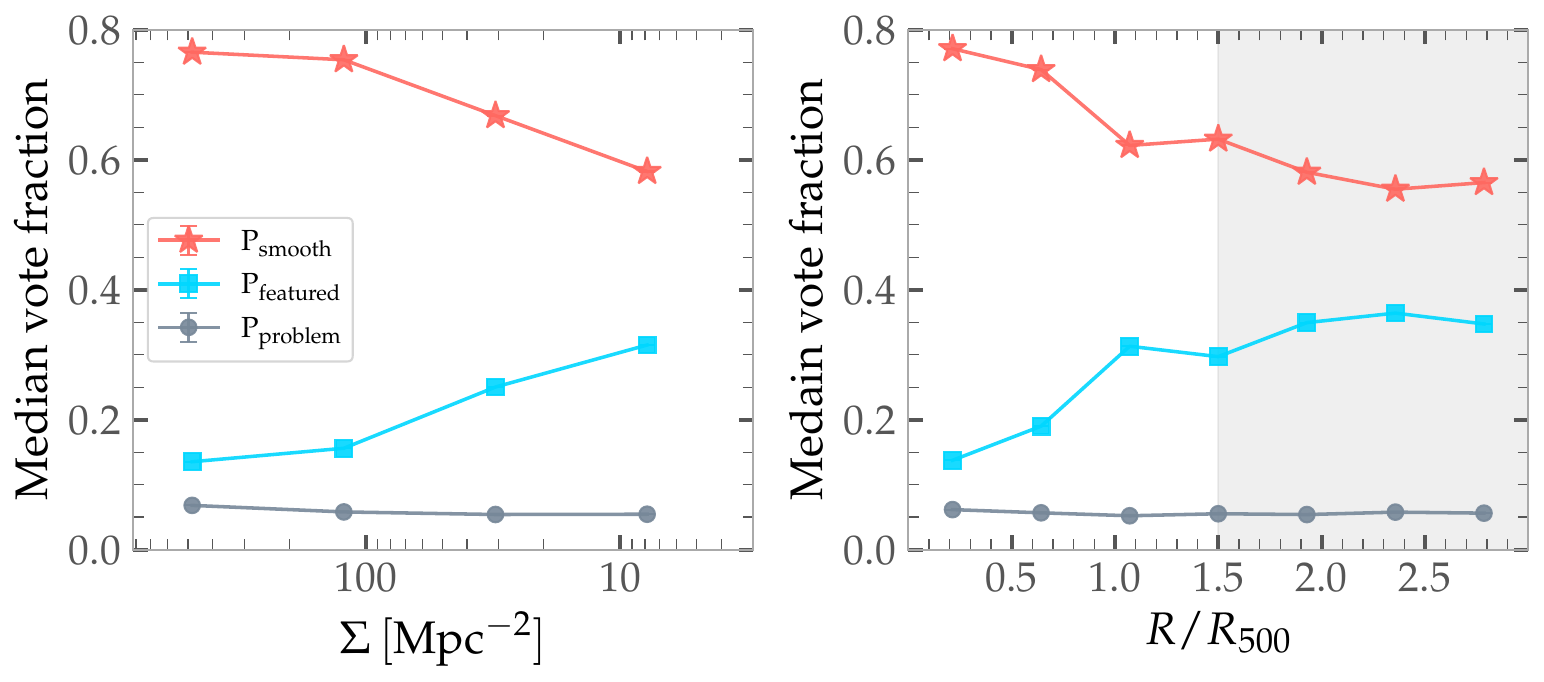}
    \caption{Median-predicted smooth (red), featured-or-disk (blue), and problem (gray) vote fractions for the cluster members as functions of binned environment indicators. \textit{Right panel}: Gray-shaded region, as in Fig.~\ref{fig:morph_fracs}.}
    \label{fig:smooth_feat_votes}
    
\end{figure}

\begin{figure}[h!]\label{LastPage}
    \begin{minipage}{0.5\linewidth}
    \includegraphics[width=.9\linewidth]{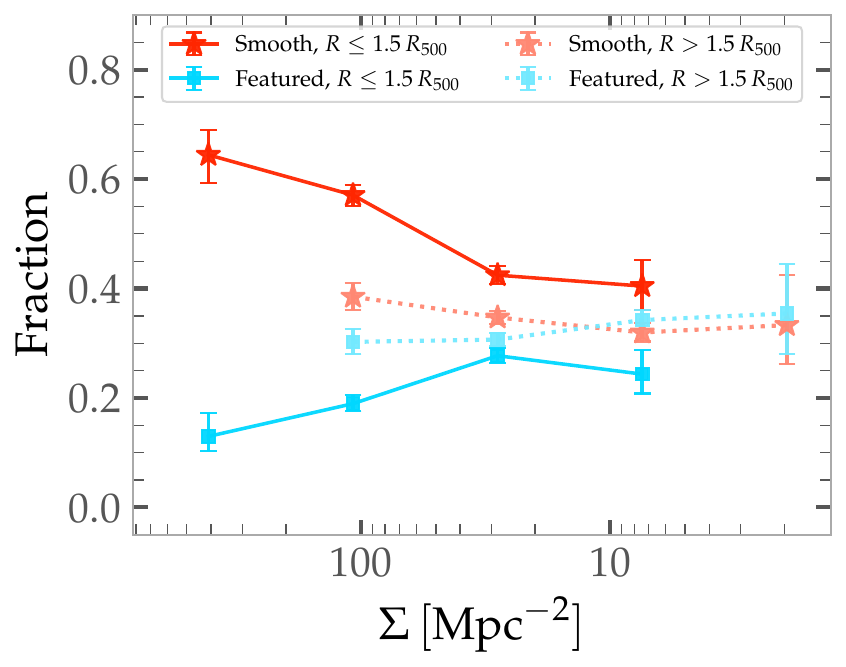}
    \caption{Same as left panel of Fig.~\ref{fig:morph_fracs} but with $\Sigma_N$ computed with $N=7$, as in \citet{cleland_ec}.}
    \label{fig:smooth_feat_dens7}
    \end{minipage}
\end{figure}

Figure~\ref{fig:smooth_feat_votes} illustrates the change of the smooth, featured-or-disk, and problem median expected vote fractions for each of the environment bins. Figure~\ref{fig:smooth_feat_dens7} shows the $T$--$\Sigma$ relation for smooth and featured-or-disk cluster members, with the local density computed using $N=7$ neighbors.

\FloatBarrier
\end{appendix}

\end{document}